\documentclass[journal]{IEEEtran}
\usepackage[utf8]{inputenc}
\usepackage{bbding}
\usepackage{microtype}
\usepackage{graphicx}
\usepackage{subfigure}
\usepackage{booktabs}
\usepackage[numbers]{natbib}
\usepackage{hyperref}

\usepackage{multirow}
\usepackage{blindtext}
\usepackage{enumitem}
\usepackage{gensymb}
\usepackage{footnote}
\usepackage{authblk}
\usepackage{tablefootnote}
\usepackage{scrextend}
\usepackage[linesnumbered,ruled]{algorithm2e}
\usepackage{amsmath,amssymb,amsfonts}
\usepackage{textcomp}
\usepackage{bm}

\newcommand{\R}[1] {\textnormal{#1}}

\newcommand{\highlighttext}[1] {#1}

\makesavenoteenv{table}
\makesavenoteenv{table*}
\makesavenoteenv{tabular}

\begin{document}
\title{RAMP-CNN: A Novel Neural Network for Enhanced Automotive Radar Object Recognition}

\author[1]{Xiangyu Gao}
\author[1]{Guanbin Xing}
\author[1]{Sumit Roy}
\author[1,2]{Hui Liu}
\affil[1]{Dept. of Electrical and Computer Engineering \\ University of Washington, Seattle, WA 98195}
\affil[2]{Silkwave Holdings, Hong Kong}
\affil[ ]{\textit{\{xygao, gxing, sroy, huiliu\}@uw.edu}}

\maketitle

\begin{abstract}
Millimeter-wave (mmW) radars are being increasingly integrated into commercial vehicles to support new advanced driver-assistance systems (ADAS) by enabling robust and high-performance object detection, localization, as well as recognition - a key component of new environmental perception. In this paper, we propose a novel radar multiple-perspectives convolutional neural network (RAMP-CNN) that extracts location and class of objects based on further processing of the {\em range-velocity-angle} (RVA) heatmap sequences. To bypass the complexity of 4D convolutional neural networks (NN), we propose to combine several lower-dimension NN models within our RAMP-CNN model that nonetheless approaches the performance upper-bound with lower complexity. The extensive experiments show that the proposed RAMP-CNN model achieves better average recall (AR) and average precision (AP) than prior works in all testing scenarios. Besides, the RAMP-CNN model is validated to work robustly under the nighttime, which enables low-cost radars as a potential substitute for pure optical \highlighttext{sensing} under severe conditions. Our code is available at \href{https://github.com/Xiangyu-Gao/Radar-multiple-perspective-object-detection}{\textit{https://github.com/Xiangyu-Gao/Radar-multiple-perspective-object-detection}}.
\end{abstract}


\section{Introduction}
\IEEEPARstart{T}{he} millimeter-wave (mmW) radars provide highly accurate object detection and localization (range, velocity and angle), largely independent of environmental conditions \cite{YONEDA2019253}. Thus, they are fast becoming indispensable in providing critical sensory inputs for environmental mapping in future autonomous vehicle operations. In challenging conditions -  nighttime, glaring sunlight, snow, rain or fog - the utility of pure optical \highlighttext{sensing} (camera and lidar) is diminished \cite{impactweather}; hence the primary objective of this paper is to enable low-cost mmW radar as a potential substitute. To achieve this, radars should deliver semantic environment perception close to what optical sensors provide.

The evolution of radar-based object recognition algorithms has been driven by recent advances in automotive radar hardware using chirp or frequency modulated continuous wave (FMCW) over 77-81 GHz RF bandwidth with integrated digital CMOS and packaging resulting in low-cost radar-on-chip system \cite{8828025}.  Texas Instrument (TI)'s state-of-art 77 GHz FMCW radar chips and evaluation boards - AWR1443, AWR1642, and AWR1843 - are built with low-power 45-nm RF CMOS process and enable unprecedented levels of integration in an extremely small form factor \cite{ti1642datasheet}. Other vendors (e.g. Uhnder) have recently unveiled a new, {\em all-digital} phase modulated continuous wave (PMCW) radar chip capable of synthesizing multiple-input and multiple-output (MIMO) radar capability with 192 virtual receivers, thereby obtaining very high angular resolution \cite{8662386}. Such high-resolution radars perform similar functions to lidars, i.e., generate dense point-cloud maps from object returns in the vicinity at a fraction of the cost of lidar systems.

\begin{figure}
    \centering
    \includegraphics[width=3.4in,trim=1 1 1 1,clip]{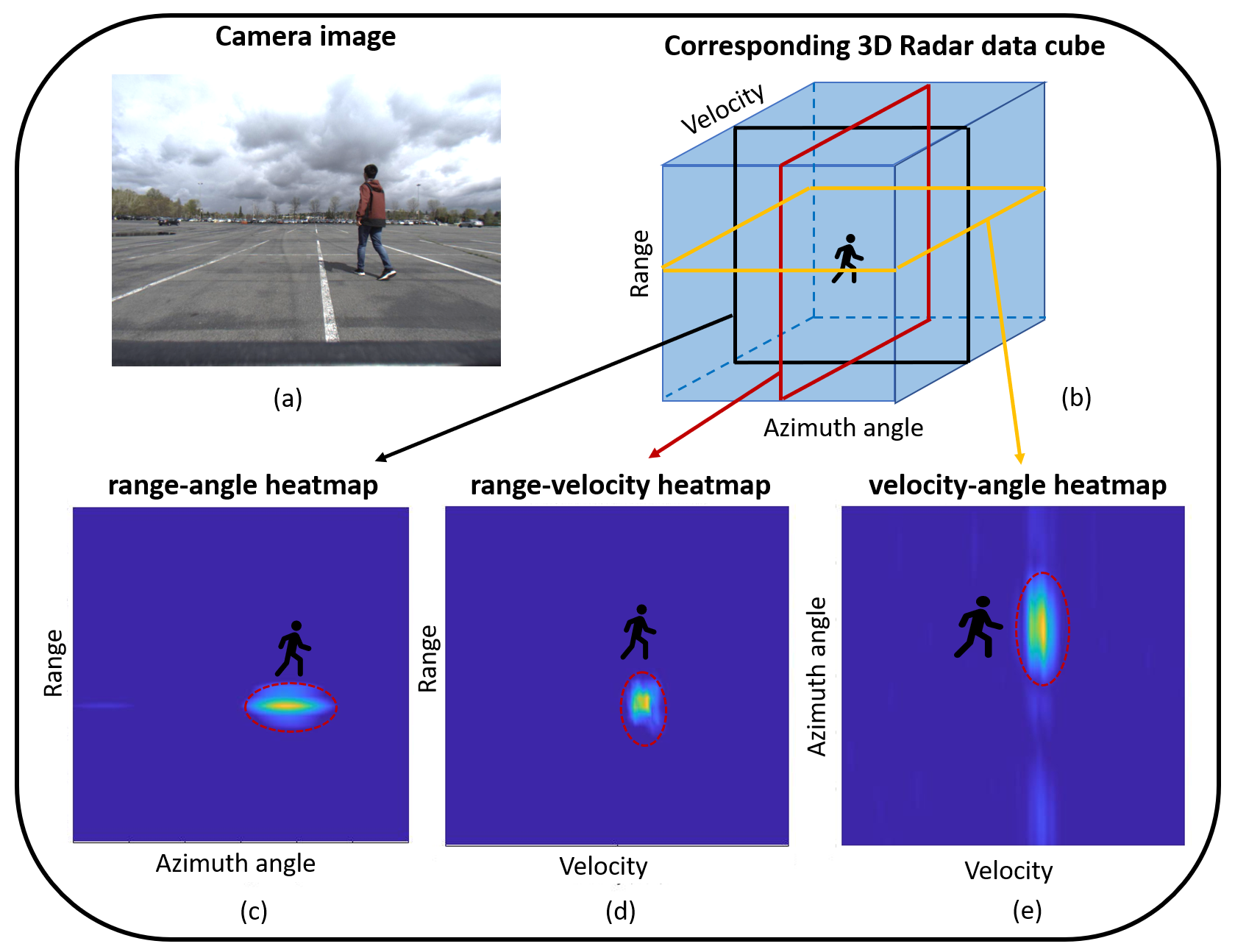}
    \caption{Abstraction of single frame input radar data and corresponding camera image: (a) The camera image of the pedestrian; (b) The range-velocity-angle radar data cube, three cross profiles of it are shown as figure (c), (d) and (e); (c) The range-azimuth angle heatmap; (d) The range-velocity heatmap; (e) The velocity-azimuth angle heatmap.}
  \label{abstraction of single framef}
  \vspace{-3mm}
\end{figure}

FMCW radars transmit a linear frequency modulated signal; the received signal reflected from a target is mixed with the transmitted signal to obtain the beat frequency, which is a function of the round trip delay and therefore can be mapped directly to range \cite{iovescu2017fundamentals}. Similarly, transmitting a train of equispaced FMCW chirps (also called a frame) allows Doppler velocity estimation for target that undergoes (relative) radial motion. Such radial motion induces a phase shift over the chirps in a range resolution cell, which is used to compute the Doppler radial velocity \cite{iovescu2017fundamentals}. Finally, the use of multiple transmitters and receivers enables azimuth localization of target by appropriate beamforming processing of the multiple transmitted waveforms reflected by the target to receiver array \cite{ti_mimo}. In summary, the analog-to-digital converted (ADC) raw radar data - has 3 dimensions: samples (fast time), chirps (slow time), and receivers) - can be mapped to the 3D radar cube with 3 new dimensions: range, Doppler velocity, and angle. In this paper, we adopt the 3-DFFT \cite{gao2019experiments} to obtain the 3D radar cube 
that is named the range-velocity-angle (\textbf{RVA}) heatmap \footnote{In general, by the heatmap we refer to the complex image resulting from the FFT operations. When for visualization purposes, we take the amplitude value of the complex heatmap.} \footnote{In this paper, angle represents the azimuth angle if not specified.}.

The small form factor of TI 77 GHz boards - while a desirable feature - limits the number of antennas that can be integrated, resulting in poor angular resolution (see Table. \ref{param}). Specifically, two targets at the same distance \highlighttext{and radial velocity} are not resolved in angle if separated by less than resolution beamwidth; even if resolvable, the spatial dimension is not well-defined. Hence to achieve reliable object recognition using such hardware, \cite{gao2019experiments, 8468324} have sought to exploit the unique movement patterns over time for different classes of objects, i.e. rely on temporal patterns over multiple frames rather than spatial discrimination from single-frame data.

\begin{table*}
\begin{center}
\caption{Parameters and configurations of TI's AWR1843 FMCW radar \cite{ti1642datasheet, gao2019experiments}}
\begin{tabular}{ll|ll}
\toprule  
Parameter & Calculation Equation& Configuration & Value\\
\midrule  
Range resolution ($R_{\R{res}}$) & $R_{\R{res}}=\frac{c_{\R{0}}}{2B}=0.23$ m  &  Frequency ($f_{\R{c}}$) & 77 GHz \\[0.75ex]

Velocity resolution ($V_{\R{res}}$) & $V_{\R{res}}=\frac{\lambda}{2N_{\R{c}} T_{\R{c}}}=0.065$ m/s \footnote{In the calculation of velocity resolution, $\lambda$ is the wavelength of transmitted signal, $T_{\R{c}}$ is the duration of one chirp, $T_{\R{c}} = \frac{1}{N_{\R{c}} f_{\R{F}}}$.} & Sweep Bandwidth ($B$) & 670 MHz \\[0.75ex]

Angle resolution ($\theta_{\R{res}}$)  & $\theta_{\R{res}}=\frac{\lambda}{N_{\R{Rx}}d\cos{\theta}} \approx 15\degree$  \footnote{In the calculation of angle resolution, $N_{\R{Rx}}$ is the number of elements of receiver array (including the MIMO virtual receiver), $d$ is the separation between receive antenna pair.} & Sweep slope ($S$) & 21 MHz/$\mu$s \\[0.75ex]

Max operating range ($R_{\R{max}}$) & $R_{\R{max}} = \frac{f_{\R{s}} c_{\R{0}}}{2S} = 28.5$ m & Sampling frequency ($f_{\R{s}}$) & 4000 Ksps\\[0.75ex]

Max operating velocity ($V_{\R{max}}$) \footnote{The operating velocity range is $\left(-V_{\R{max}}, V_{\R{max}}\right)$. Similarly, the operating angle range is $\left(-\theta_{\R{max}}, \theta_{\R{max}}\right)$} & $V_{\R{max}} = \frac{\lambda}{4T_{\R{c}}} = 8.3$ m/s & Num of chirps in one frame ($N_{\R{c}}$) & 255\\[0.75ex]

Max operating angle ($\theta_{\R{max}}$) & $\theta_{\R{max}} = \sin^{-1}{\left(\frac{\lambda}{2d}\right)} = 90\degree$ & Num of samples of one chirp ($N_{\R{s}}$) & 128 \\[0.75ex]

 & & Num of transmitters, receivers & 2, 4\\[0.75ex]
 
 & & Frame rate ($f_{\R{F}}$) & 30 FPS\\
\bottomrule 
\label{param}
\end{tabular}
\end{center}
\vspace{-6mm}
\end{table*}

Traditional radar object recognition algorithms are based on the statistical signal processing and manual feature selection \cite{5494432, 2012AdRS...10...45B, Scheiner_2018}. For example, it is usual to apply the constant false alarm rate (CFAR) \cite{doi:10.1036/0071444742} algorithm and DBSCAN clustering \cite{Ester:1996:DAD:3001460.3001507} algorithm to detect the location of objects. Then predefined features of the detected objects, such as SNR, range profile, Doppler spread \cite{5494432}, number of detections, and spatial distance \cite{Scheiner_2018} are extracted and combined to determine the classes of objects. Recent advances in deep learning (DL) have promoted novel approaches for automating feature selection. For general DL methods, the algorithmic time complexity can be reduced enormously by implementing suitable pre-processing on the input data. However, too much pre-processing risks losing key information embedded in the raw data; \highlighttext{we therefore} seek the right trade-off between performance and efficiency.

While several prior works \cite{gao2019experiments,8468324, Major_2019_ICCV, 8890199, wang2020rodnet} explore radar object recognition with various input data formats using NN, none has ever combined the spatial and temporal domain information, i.e., by jointly processing the 3D radar cube {\em sequences} (from multiple frames). Our fundamental contribution is a {\em deep learning} network design with 3D radar cube sequences as input that approaches performance upper bound by exploiting all available information (Section \ref{physical significance}). 

However, it is impractical to implement 4D (3D from radar cube plus time sequences) convolution processing as the resulting computation complexity is unacceptable for real-time perception. Therefore, we propose to combine several lower-dimension (3D) models, which nonetheless exceeds the performance of prior methods with acceptable computation complexity (see Table. \ref{time complexity} and \ref{space complexity}). Basically, each 3D radar cube (RVA heatmap) is sliced into 2D images from 3 perspectives, that is, range-angle (RA) heatmap, range-velocity (RV) heatmap, and velocity-angle (VA) heatmap. The RA, RV, and VA heatmap sequences are then processed by three parallel DL models to generate different feature bases, which are fused to make the object recognition decision (see Fig. \ref{network}). We name above radar network architecture RAMP-CNN.

\begin{figure}[h]
    \hspace{-1em}
    \vspace{-3mm}
    \includegraphics[width=3.8in,trim=1 5 1 1,clip]{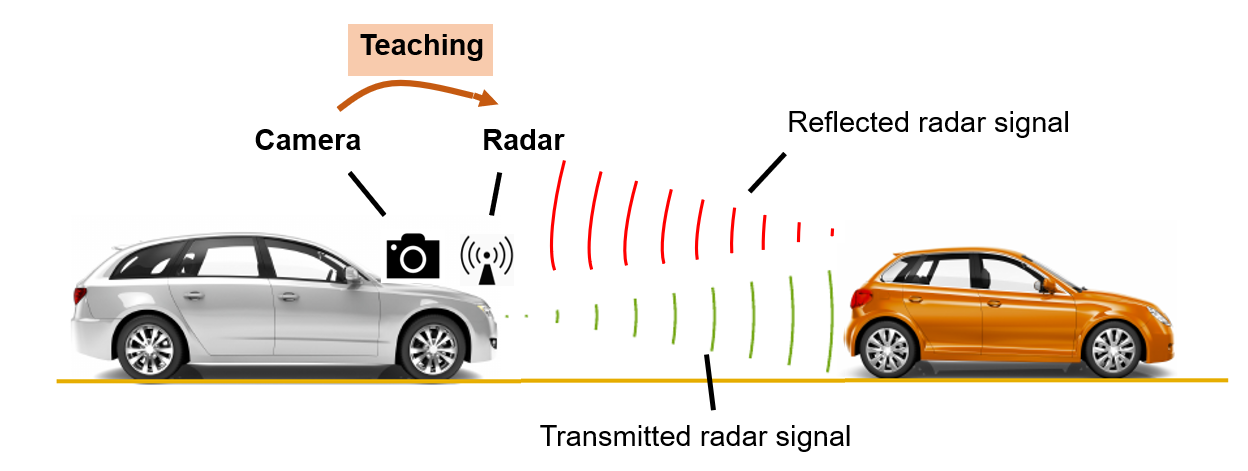}
    \caption{The vehicular radar-camera system \protect \footnotemark}
  \label{cam_rad_sys}
  \vspace{-3mm}
\end{figure}

Supervised learning methods need a huge amount of training data and corresponding ground truth labels, a challenge for every DL-based approach. In particular, human labeling of radar data is unreliable even in good conditions, in contrast to labeling camera images. To solve this problem, we propose to set up a vehicular radar-camera system as shown in Fig. \ref{cam_rad_sys} and \ref{platform} to collect the synchronized radar data and camera images for building the UWCR dataset. Cameras are used for teaching radars the locations (range and angle) and classes of objects under good light and weather conditions, which is achieved by implementing the object detection and depth estimation algorithm on captured images (Section \ref{cam sup}).

\footnotetext{Fig. \ref{cam_rad_sys} is modified from \cite{aacc_matlab} and the abstract figure is modified from \cite{ADASradar}.}

To further improve the performance of RAMP-CNN model, we propose the following two modifications to avoid overfitting in the training stage, which have been validated in Ablation study (Section \ref{ablation study}).

\subsubsection*{Radar data augmentation algorithms}

Data augmentation encompasses a suite of techniques that enhance the size and quality of training dataset such that better DL models can be built. Traditional image augmentation algorithms include geometric transformations, color space augmentations, mixing images, etc. In Section \ref{radar_aug}, we propose 4 basic data augmentation operations - flipping, translating, interpolating, and mixing - that work for radar data by accounting for radar imaging physics: energy loss with range and nonuniform angular resolution\footnotemark[13].

\subsubsection*{New loss function design}
The feature basis from the RA, RV, and VA input - named RA, RV, and VA features respectively - are fused within the feature fusion module. When fusing, RA features remain unchanged while RV and VA features are mapped to the range-angle domain. The resulting issue is that NN may well give more weights to the straightforward and accessible RA features than other velocity-based features, leading to overfitting. Therefore, to push NN to 
effectively utilize RV and VA features, we add a new term - that only takes RV and VA inputs - to original loss function (Section \ref{loss design}). 

In summary, the main contributions of this paper are four-fold:
\begin{itemize}
\item Design a novel temporal-spatial RAMP-CNN model that jointly processes the 3D radar cube sequences to achieve superior performance than all prior works, as validated by extensive testing with our UWCR dataset.
\item \highlighttext{The proposed RAMP-CNN model is validated to work robustly under the nighttime, where cameras (and other passive optical sensors) are largely ineffective.}
\item For training the RAMP-CNN model, we propose and establish a vehicular radar-camera system that uses cameras to teach radars the locations and classes of objects under the good light and weather conditions.
\item To avoid overfitting in training stage, we propose the modified data augmentation algorithms suitable for radar data and design a new loss function that pushes the RAMP-CNN model to utilize more velocity-related features.
\end{itemize}

The rest of this paper is organized as follows. Several relevant prior works are commented in \highlighttext{Section \ref{related works}}. The principle of radar preprocessing algorithm is introduced in \highlighttext{Section \ref{radar data preprocessing}}. The RAMP-CNN architecture and radar data augmentation algorithms are presented in \highlighttext{Section \ref{RAMP-CNN model architecture}} and \ref{radar_aug}. We describe the experiment details including the evaluation results in \highlighttext{Section \ref{experiments}} and analyze the RAMP-CNN model in \highlighttext{Section \ref{Analysis}}. In the end, we conclude the paper and propose future work. 

\section{Related Work \label{related works}}
We comment on the relevant prior works \cite{gao2019experiments, 8468324, Major_2019_ICCV, 8890199, wang2020rodnet, 9046713} that have attempted radar object classification with various radar data input formats.

\cite{8468324} uses the short-time Fourier transform (STFT) intensity (heatmap) as the input and implements several existing DL models to extract micro-Doppler patterns from it, with up to $93\%$ recognition accuracy when evaluated for three classes discrimination: car, pedestrian and cyclist. Further, \cite{gao2019experiments} proposes a new framework to pre-process the raw radar data and enhance radar object classification by incorporating not only the micro-Doppler pattern but also the spatial information. However, the above methods are two-stages architectures where the regions of interest (i.e., the locations of objects) need to be found before the classification, which usually takes longer inference time than one-stage methods. Besides, in \cite{gao2019experiments, 8468324}, there are several prepossessing procedures (i.e., CFAR detection, DBSCAN clustering, etc) before feeding the radar input to NN, which may make the information incomplete.

\cite{8890199} presents a single shot detection and classification system in urban automotive scenarios, which is based on the YOLO system applied to the pre-processed range-Doppler-angle power spectrum with 77 GHz FMCW radar. To feed the 3D radar power spectrum into 2D YOLO network, \cite{8890199} condenses the angular domain by choosing the maximum. The range-angle domain is the main perspective to observe the objects' reflection ability and shapes such that condensing angular domain is not the best choice. Similarly, \cite{9046713} also takes the \highlighttext{range-Doppler} radar data  as input and predicts object class with a CNN.

\cite{Major_2019_ICCV} illustrates a DL-based vehicle detection solution that operates on the absolute-valued range-velocity-angle radar tensor. The ability of accuracy vehicle detection in high way scenario mainly relies on recognizing the energy distribution on the range and angle dimension (i.e., the contribution of Doppler dimension to detection is small). This may be hard for solving the various object recognition problem we have. 

\cite{wang2020rodnet} shows a radio object detection network (RODNet) to detect objects purely from the processed radar data in the format of range-angle heatmap sequences. \cite{wang2020rodnet} exploits the temporal information behind the change of spatial patterns across frames. However, for each frame, \cite{wang2020rodnet} just randomly picks one range-angle heatmap of a chirp signal, which is convenient but gives up the abundant velocity information behind the phase change across chirps.

\highlighttext{Besides object classification, a bunch of NN-based radar applications have been attempted, such as human activity classification \cite{8822709, 8716728}, hand-gesture recognition \cite{8610109}, and the armed and unarmed personnel recognition \cite{8626156}.}

\section{radar Data Preprocessing \label{radar data preprocessing}}

\begin{figure}[h]
    \hspace{-0.5em}
    \includegraphics[width=3.8in]{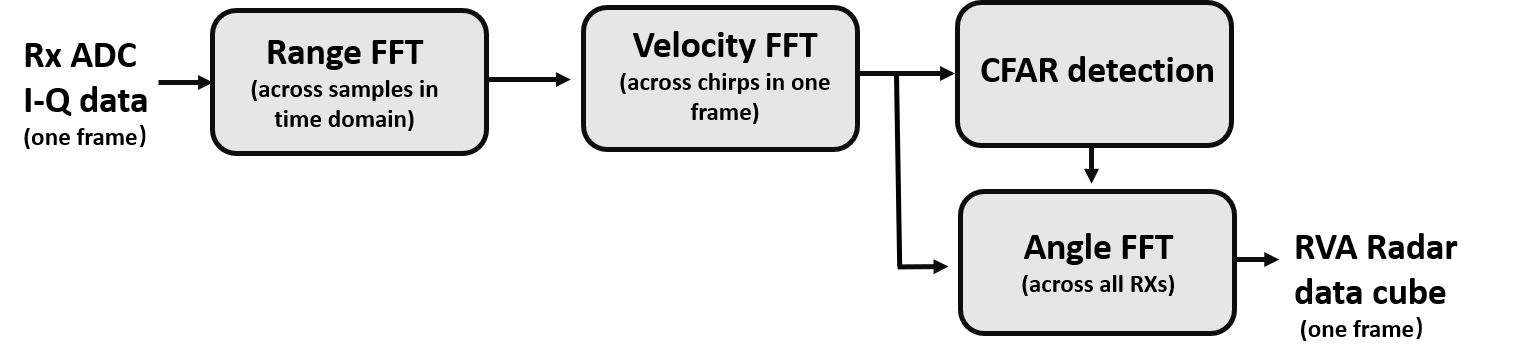}
    \caption{Basic radar signal processing chain}
    \label{workf}
    \vspace{-3mm}
\end{figure}

An FMCW radar transmits a train of FMCW chirp signals - named a frame - and then mixes the received echo with the local reference (transmitted signal) to yield the intermediate frequency (IF) signal. The IF signal is digitized by Quadrature ADC and then processed by the 3-DFFT algorithm shown in Fig.\ref{workf}. The 3-DFFT algorithm consists of 3 discrete fast Fourier transforms (DFFT) that estimate the spectrum of range, Doppler velocity, and angle respectively. \highlighttext{The third FFT (Angle FFT) is performed across receivers at every cell of the range-velocity spectrum (Velocity FFT output). Before that, we implement the CFAR \cite{doi:10.1036/0071444742} algorithm on range-velocity spectrum and then compensate the Doppler-induced phase change \cite{8052088} at the locations where CFAR produced detections.}

\subsection{Range Estimation \label{range part}}

For the target at range $r$, the resulting beat signal has a frequency $f_{\R{b}}=\frac{S2r}{c_{\R{0}}}$ in the IF band, where $S$ is the slope of a chirp signal, and $c_{\R{0}}$ is speed of light. To estimate the beat frequency, a fast Fourier transform ({\em Range FFT}) is used to convert the time domain IF signal into the frequency domain; the peaks in the resulting spectrum is used to detect resolved objects.
The resolution of FFT-based range estimation is determined by the swept RF bandwidth $B$ of the FMCW system \cite{doi:10.1036/0071444742}, i.e., $R_{\R{res}}=\frac{c_{\R{0}}}{2B}$. In our experiments with the TI system, the FMCW signal is configured for 670 MHz swept bandwidth, and the expected range resolution is 0.23 m.

\subsection{Velocity Estimation}

\begin{figure}[h]
    \centering
    \includegraphics[width=3in]{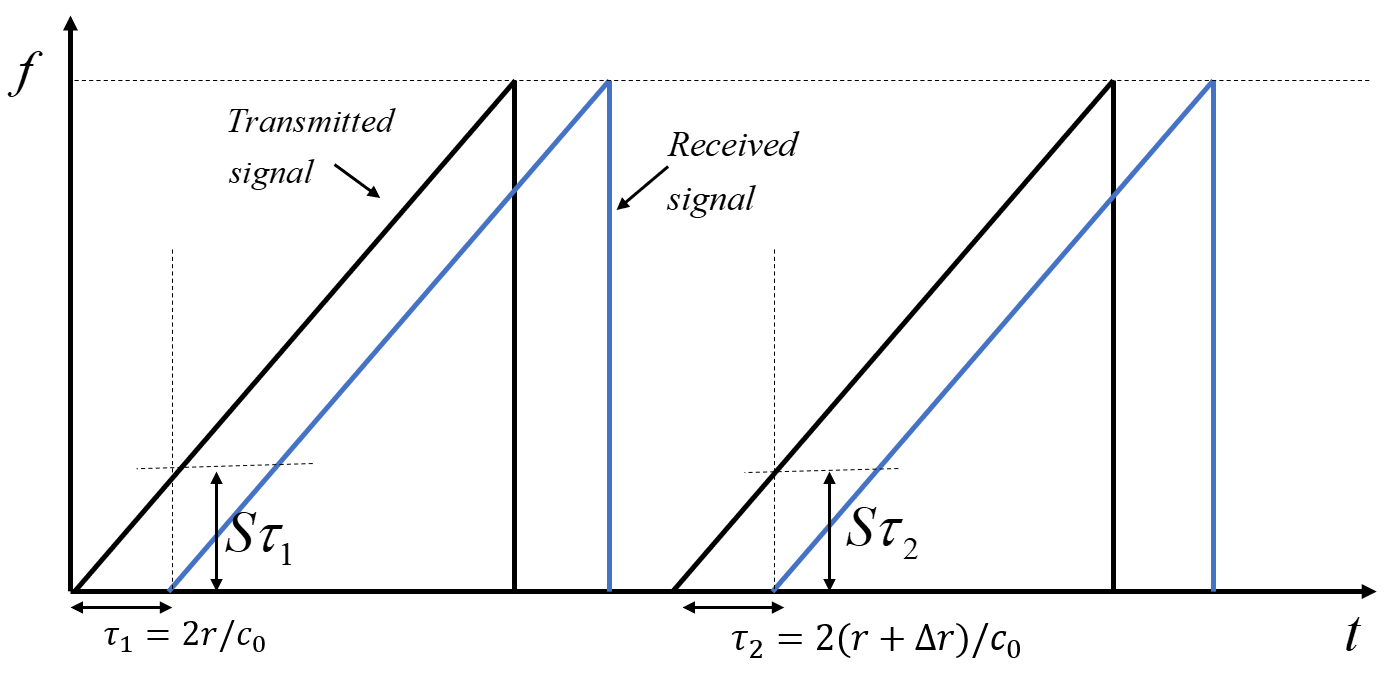}
    \caption{The illustrations of the range and velocity measurement}
    \label{fig1}
    \vspace{-3mm}
\end{figure}

Any radial object motion $\Delta r$ (shown in Fig. \ref{fig1}) relative to the radar between consecutive chirps will cause a frequency shift $\Delta f_{\R{b}} = \frac{2S\Delta r }{c_{\R{0}}}$ as well as a phase shift $\Delta \phi_v = 2\pi f_{\R{c}}\frac{2\Delta r}{c_{\R{0}}}=\frac{4\pi v T_{\R{c}}}{\lambda}$ in the beat signal \cite{gao2019experiments, iovescu2017fundamentals}, where $f_{\R{c}}$ is the carrier frequency, $v$ is the object velocity, $T_{\R{c}}$ is the chirp period, and $\lambda$ is the wavelength. Compared to the beat frequency shift, the phase shift is more sensitive to the object movement \cite{iovescu2017fundamentals}. Hence, it is common to execute a fast Fourier transform ({\em Velocity FFT}) across the chirps to estimate phase shift and then transform it to velocity estimation. The velocity resolution of this method is given by: $V_{\R{res}}=\frac{\lambda}{2N_{\R{c}} T_{\R{c}}}$ \cite{iovescu2017fundamentals}, where $N_{\R{c}}$ is the number of chirps in one frame. The expected velocity resolution is 0.065 m/s given the configuration $N_{\R{c}}=255$, and $T_{\R{c}}=120 \, \mu\text{s}$.
 
\subsection{Angle Estimation}

\highlighttext{Angle estimation is conducted via processing the signal at a receiver array composed of multiple elements. The return from a target located at far field and angle $\theta$ results in the steering vector $\bm{a}_{\R{ULA}}(\theta)=[1, e^{-j2\pi d\sin{\theta}/\lambda}, \cdots, e^{-j2\pi (N_{\R{Rx}}-1)d\sin{\theta}/\lambda}]^{\R{T}}$ as} \highlighttext{the uniform linear array output \cite{526899}, where $d$ denotes the inter-element distance. Hence a fast Fourier transform across Rx elements ({\em Angle FFT}) can easily resolve objects according to their arrival angles $\theta$ \cite{gao2019experiments, ti_mimo}.}

\highlighttext{TI chips \cite{ti1843evm} provide the MIMO radar capability that forms a larger virtual array with orthogonal transmit waveforms \cite{mimophdthesis}, which also enables a greater degree of freedom capability and better angle resolution \cite{ti_mimo}. We adopted the TDM-MIMO \cite{ti_mimo, 7060251} configuration with 2 Tx and 4 Rx for all collected data such that the resulting virtual array consists of 8 elements.}

\highlighttext{For non-stationary targets, the motion-induced phase errors should be compensated on the virtual antennas (elements corresponding to the second Tx in case of TDM-MIMO) before the Angle FFT. According to \cite{8052088}, these virtual elements are corrected via rotating phase by $\frac{\Delta \phi_v}{2}$, half of the estimated Doppler phase shift, where $v$ is obtained from the CFAR detection results on the range-velocity spectrum.}

\highlighttext{The angle resolution for FFT processing is known to be  $\theta_{\R{res}}=\frac{\lambda}{N_{\R{Rx}}d\cos{\theta}}$ \cite{iovescu2017fundamentals}. For boresight $\theta = 0^{\circ}$, $N_{\R{Rx}}=8$, and $d=\frac{\lambda}{2}$, the angle resolution is approximated to $15^{\circ}$.}

\begin{figure*}[h]
\centering
\resizebox{\textwidth}{!}{\includegraphics[width=6.2in]{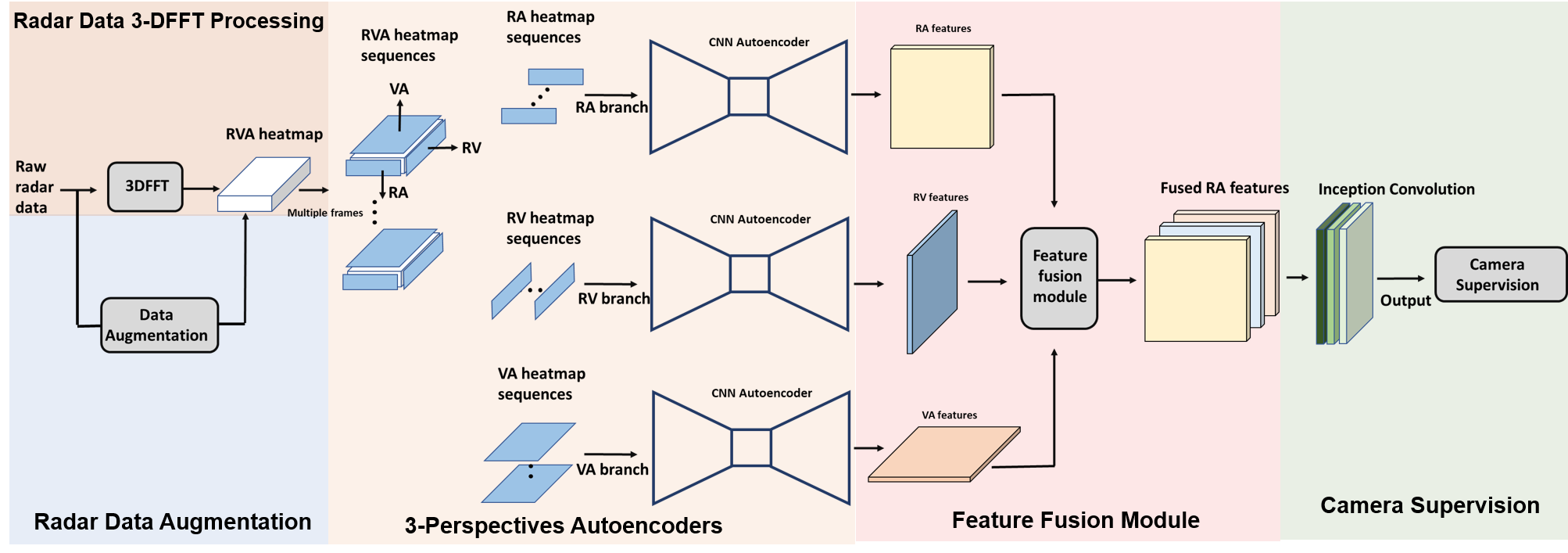}}
\caption{The architecture of RAMP-CNN model}
\label{network}
\vspace{-3mm}
\end{figure*}

\begin{figure*}
    \centering
    \includegraphics[width=6in,trim=1 1 1 1,clip]{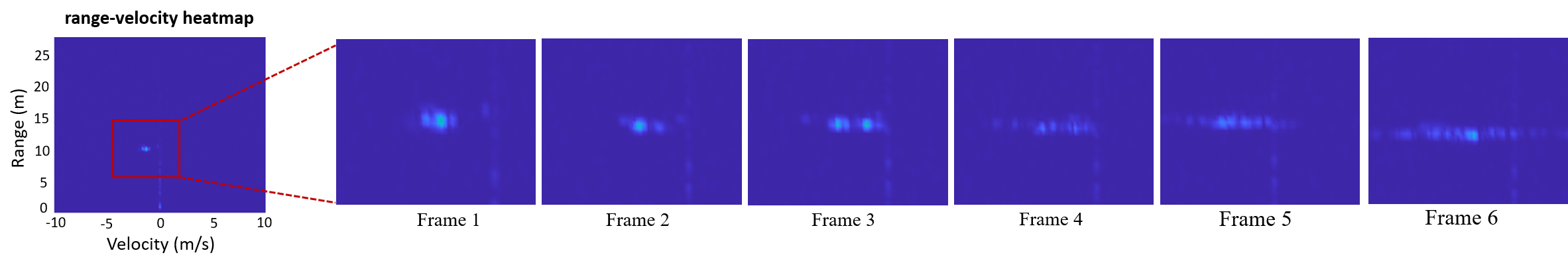}
    \caption{Visualization of a pedestrian's movement patterns: we zoom in partial regions of 6 range-velocity heatmaps of a moving pedestrian. The x-axis is velocity and the y-axis is the range. We can observe that this pedestrian has a small location movement over 6 frames but a big change of velocity patterns.}
  \label{microdop}
  \vspace{-3mm}
\end{figure*}

\section{RAMP-CNN model: A Convolutional Neural Network for radar Data \label{RAMP-CNN model architecture}}

\subsection{3-Perspectives Autoencoders Design}

\highlighttext{As shown in Fig. \ref{network}, the main body of the RAMP-CNN architecture is composed of 3 convolutional autoencoders. These autoencoders extract features from the heatmap sequences of different perspectives - that is, RA, RV, and VA respectively.}

\subsubsection*{\textbf{Why Convolutional Autoencoder}}

 CNNs are known for excellent feature extractor in some tasks, e.g, object detection, and segmentation. The convolutional autoencoders (CAE) - consisting of an encoder and a decoder - render a compact feature representation of the input, by learning the optimal filters that minimize the reconstruction error \cite{Shou_2017_CVPR, stackautoencoder}. The output feature representation/basis is the same size grid as input and the cell value of grid is the feature strength. Each (input) image-plane pixel location maps to multiple feature grid indices. Thus, operations such as weighted-sum followed by a suitable non-linearity on the feature grid cells can be used to determine the presence of particular-type objects at one location. The parameters for such operations (i.e., weights and bias) can be trained by suitable NN iteration.
 
\subsubsection*{\textbf{The Physical Significance of Network Design \label{physical significance}}}
The first CAE processes the \highlighttext{complex-valued} RA heatmap sequences with 3D {\em conv} layers and {\em transposed conv} layers. \highlighttext{Similar to \cite{wang2020rodnet}, we pick one RA heatmap from each frame to form the} \highlighttext{heatmap sequences, and the singled out RA heatmap is obtained by computing Range FFT and Angle FFT at an arbitrarily selected chirp \footnote{For the range bin where there exists CFAR detections, we pick its maximum-intensity velocity and use it to compensate the Doppler-induced phase error on virtual receiver elements before Angle FFT.}.} For the RA heatmap sequences input, those 3D convolution operations take advantage of not only the object's spatial patterns in a single frame but also the temporal information behind the change of spatial patterns across frames. Some aspects of spatial patterns - like the distribution of reflection intensity - directly contribute to object recognition, e.g., larger objects (vehicles) contain more stronger-reflectors than small objects (humans). 

As the RA heatmap input is with the complex-valued format, the temporal change of spatial patterns across multiple frames can be expressed as the change in both amplitude and phase. Particularly, the phase change of mmW signal along time is sensitive to the object movement, e.g. 1mm position movement results in phase shift $\Delta\phi=\pi$ for 77 GHz radar. While the sampling rate of RA heatmap input is a bit lower (30 FPS), we still believe the embedded phase shift would provide additional benefit compared to the amplitude-only input.

The second and third CAE process the absolute-valued RV and VA heatmap sequences respectively \footnote{We adopt the absolute-valued RV and VA heatmap here, since the phase change we are interested in have been preprocessed with Velocity FFT and been represented in the Doppler domain.}. The RV and VA heatmaps are calculated from the original RVA heatmap by summing the power over the omitted dimension. What two CAEs have in common is: in single RV or VA heatmap, they extract features from the distribution of range-velocity or velocity-angle cells; while across multiple heatmaps, they extract object’s movement patterns from the change of radial velocity with time. These two CAEs essentially utilize the abundant velocity-based information behind the phase change across chirps within each frame, which is the biggest difference from \cite{wang2020rodnet}.

To illustrate, different classes of moving objects exhibit different movement patterns. From Fig. \ref{microdop}, we can visualize the movement patterns of a pedestrian as the change of radial velocity with time. The velocity versus time relationship is also known as the STFT heatmap \cite{gao2019experiments} that highlights specific micro-Doppler signatures of human gait.

\begin{figure}[h]
    \centering
    \includegraphics[width=3.5in]{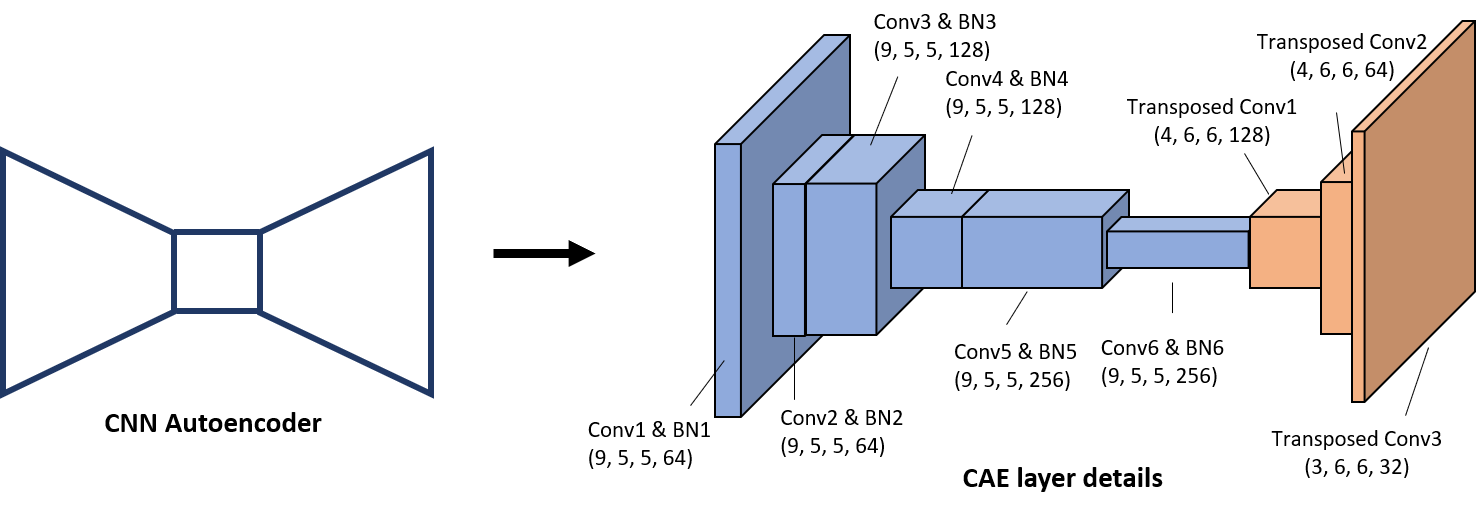}
    \caption{Details of Convolutional AutoEncoder (CAE), consist of six 3D {\em conv} layers and three 3D {\em transposed conv} layers.}
  \label{cdc}
  \vspace{-3mm}
\end{figure}

\subsubsection*{\textbf{Network Details}}
 We adopt the 3D Convolutional-De-Convolutional \cite{Shou_2017_CVPR} (shown in Fig. \ref{cdc}) model as our CAE, which is effective in summarizing spatio-temporal patterns from raw data into high-level semantics.
 
Each CAE includes six 3D {\em conv} layers and three 3D {\em transposed conv} layers. All 3D {\em conv} layers are followed by a {\em bn} (batch-normalization) layer and the {\em ReLU} activation function. The first two 3D {\em transposed conv} layers are followed by the {\em PReLU} activation function. The layer details (including parameter selection) of CAE are presented in Fig. \ref{cdc}. For illustration, the first blue cuboid part in Fig. \ref{cdc} represents the 3D {\em conv} layer and a {\em bn} layer. The kernel size of the 3D {\em conv} layer is (9, 5, 5) and the number of output feature channels is 64. 

To preserve the phase information in RA heatmap input, we represent complex-valued heatmap by two real-valued channels that store the real part and imaginary part respectively in the first CAE following \cite{8578866}. While for RV and VA heatmap inputs, it suffices to only keep the absolute value and use the one-channel representation.

\subsection{Feature Fusion Module \label{fusionsec}}
The feature basis extracted from RA, RV, VA inputs all support the final classification decision. Our final output is expressed as an image in RA domain, which means the RA feature can be directly inputted to the network to obtain the corresponding RA-format output. The key issue is how to further use RV and VA features to support an improved final classification. This is similar to initial human perception using the visual sensor (eyes) supported by supplementary sense organs (ears, nose) for final determination. A person with impaired eyesight will rely more on other sensors, e.g. acquire initial angle information/feature via the ear \footnote{The ear does not provide good range localization, and hence suggests an equal probability prior to range.}.

\begin{figure}
    \hspace{-2.em}
    \includegraphics[width=3.7in,trim=1 1 1 1,clip]{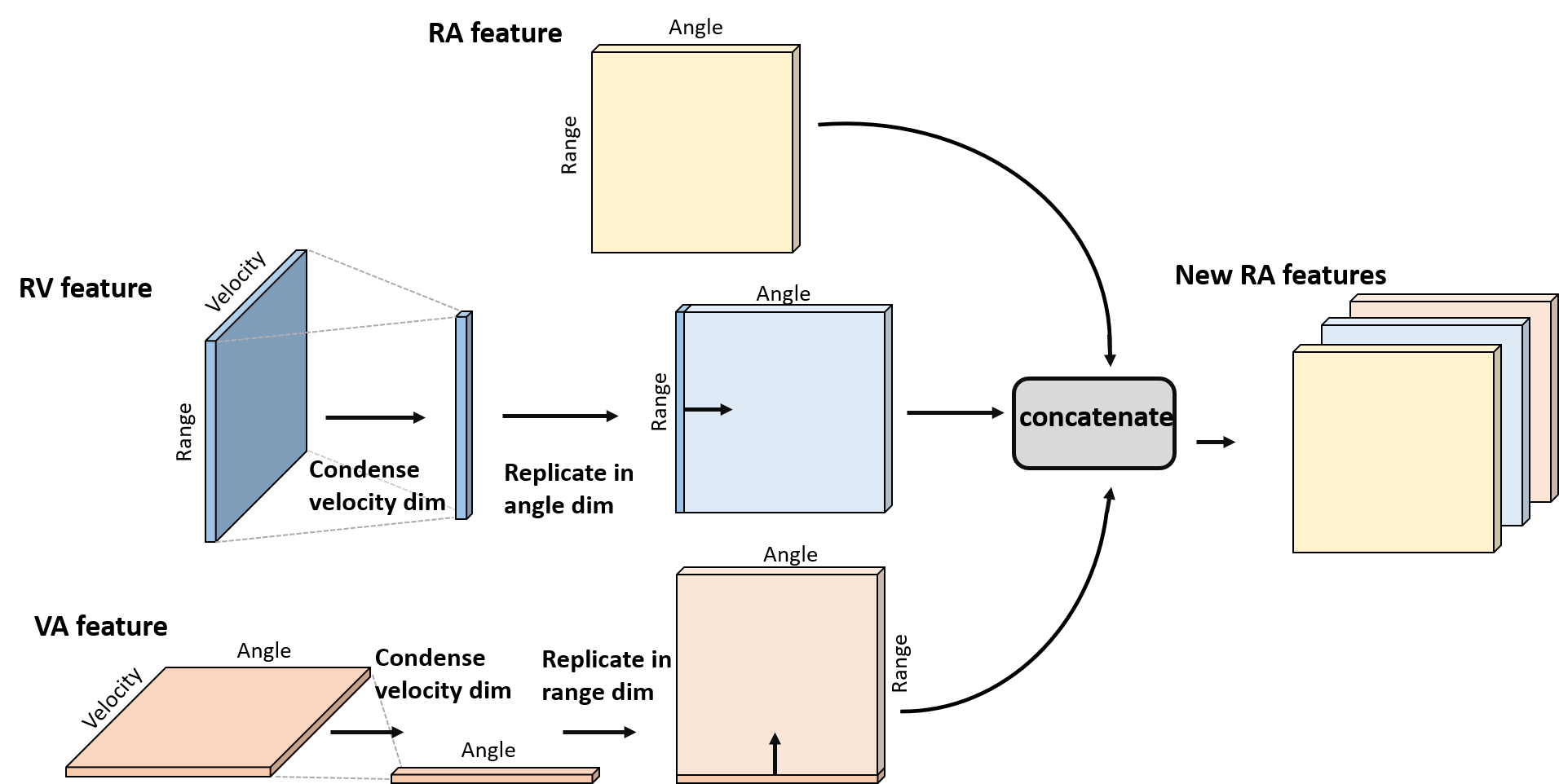}
    \caption{The structure of feature fusion module}
  \label{feature}
  \vspace{-3mm}
\end{figure}

The above analogy applied to radar processing suggests how to use the VA feature - that provides good azimuth angle information and no range information. As shown in Fig. \ref{feature}, VA feature is condensed along the velocity dimension by summing, then the condensed vector is replicated in the missing dimension - range. Similarly, we replicate the RV feature along the angle dimension. Thereafter, we concatenate all features along the channel dimension and input them to the deep network for classification decisions.

\subsubsection*{Convolution Layers after Feature Fusion Module}

There are two {\em conv} layers that take as input the fused features and make recognition decisions: one 3D {\em inception} layer, and one ordinary 3D {\em conv} layers with kernel size (3, 3, 3). Note that the ordinary {\em conv} layer operates on time, range and angle dimension, while the {\em inception} layer operates on channel, range and angle dimension of fused features. To avoid collapsing the time dimension on {\em inception} layer, we repeat the operation on each timestamp and concatenate all inception results along time dimension. 

The 3D {\em inception} layer includes 3 convolution kernels: (3, 5, 5), (3, 3, 3), (3, 1, 21). The first two kernels allow NN to take advantage of multi-level feature extraction, i.e. it extracts both general $5 \times 5$ size feature and local $3 \times 3$ size features. The last kernel with dilation 6 is used to push the NN to observe a larger area in angle - hence to solve the false alarm problem on the side-lobes \footnote{The side-lobe in radar heatmap is easy to be recognized as objects of a certain class. This is because the convolution-kernel operators of CAEs do not force each feature to be global (i.e., to span the entire visual field) \cite{stackautoencoder}.}. We make it dilated convolution - with angular kernel size 21 and dilation 6 - to cover almost all angle cells, as well as to reduce complexity.

\subsection{Camera Supervision \label{cam sup}}
\begin{figure}[h]
    \centering
    \includegraphics[width=4in]{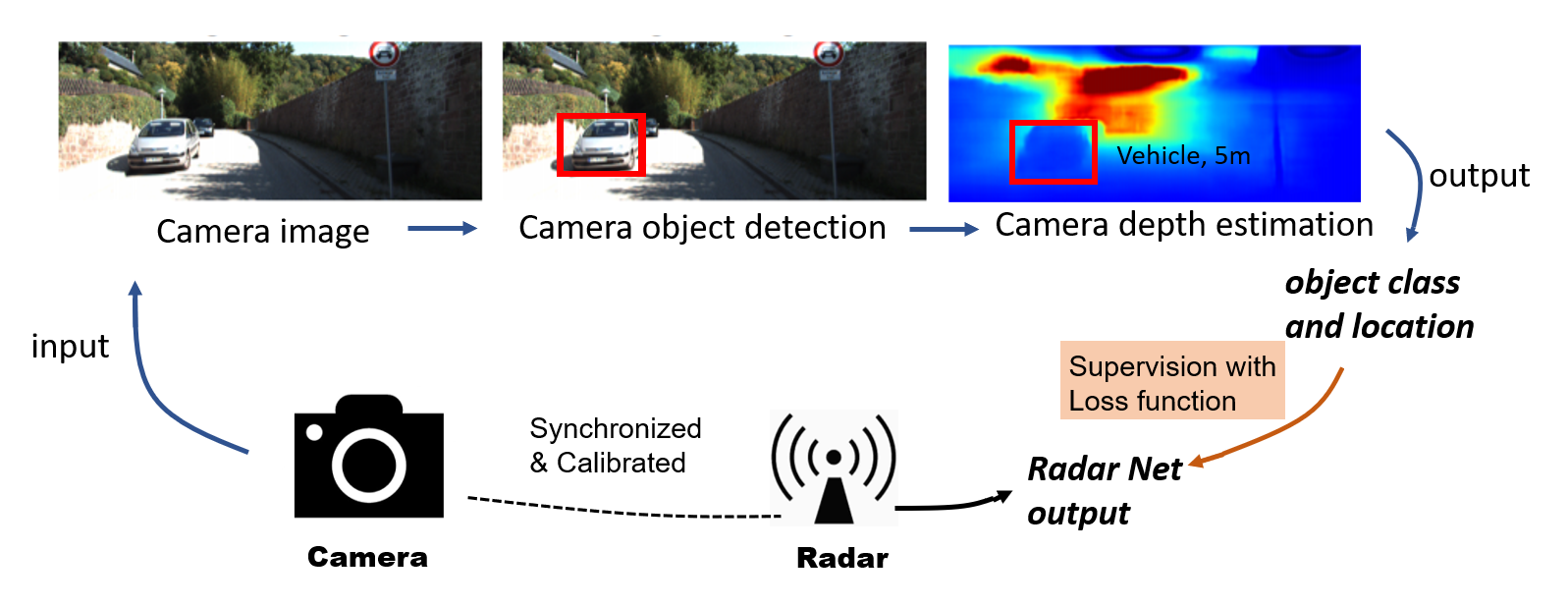}
    \caption{The framework of camera teaching radar for the training stage}
  \label{cam_tea_rad}
  \vspace{-3mm}
\end{figure}

The established radar-camera system shown in Fig. \ref{platform} generates synchronized camera images and raw radar data. As shown in Fig. \ref{cam_tea_rad}, we apply the object detection \cite{he2017mask} and depth estimation \cite{monodepth17, 10.1145/3343031.3350924}  algorithm on captured camera images to obtain the locations (i.e., range and azimuth angle) and classes of all objects of interest in the scene. Under the good light and weather conditions, the obtained information from cameras is treated as ground truth for supervising the output of RAMP-CNN. Note that camera assisted radar learning only happens in the training stage, while in testing, radar acts independently.

To ease the training burden, we use the center keypoint to represent the existence of objects following \cite{law2018cornernet}. For each ground truth center point $\mathbf{p}$ with location $(p_r, p_\theta)$, class id $p_c$ and frame id $p_t$, we compute its Gaussian representation:
\begin{equation}
    Y_{t, r,\theta,c} =
    \begin{cases}
    \exp(-\frac{(r-p_r)^2 + (\theta-p_\theta)^2}{2\sigma_p^2})  & \text{if} \; c = p_c \; \text{and} \; t = p_t\\
    0 & \text{otherwise} \\
    \end{cases}
\end{equation}

\noindent where $\sigma_p$ is an object size-adaptive standard deviation. We then splat all ground truth center points onto $Y\in [0,1]^{D \times  W\times H \times C}$, and take the element-wise maximum if two Gaussians of the same class and same frame overlap \footnote{The symbols $D$, $W$, $H$ and $C$ here represent the size of $Y$ on time, range, azimuth angle and class dimension respectively.}. $Y$ is used as the ground truth in loss function.

\subsection{Loss Function for All-perspectives Learning \label{loss design}}

Let $X_{\R{RA}}, X_{\R{RV}}, X_{\R{VA}}$ be the input RA, RV and VA heatmap sequences, the aim of RAMP-CNN model is to predict center-point heatmap sequences $\hat{Y}\in [0,1]^{D \times W \times H \times C}$ in RA domain, where $\hat{Y}_{t, r,\theta,c} = 1$ corresponds to a detected center point at range $r$, azimuth angle $\theta$, frame $t$ and class $c$, while $\hat{Y}_{t,r,\theta,c} = 0$ represents background. \highlighttext{The prediction $\hat{Y}$ includes a map for every frame time.} The center point types of each map include $C=3$ classes of objects: pedestrian, cyclist, and car.

For the prediction $\hat{Y}$ and ground truth $Y$, the training objective is a modified penalty-reduced pixelwise logistic regression with focal loss \cite{zhou2019objects, lin2017focal}:
\begin{equation}
\resizebox{0.48 \textwidth}{!}{
    $\begin{aligned}
        L_{\hat{Y}Y} =& \frac{-1}{N_{\R{obj}}}\sum_{t}  \sum_{r} \sum_\theta \sum_c\\
        &\begin{cases}
            \kappa(1-\hat{Y}_{t,r,\theta,c})^\alpha \log(\hat{Y}_{t,r,\theta,c}) & \text{if} \; Y_{t,r,\theta,c} = 1 \\
            \kappa(1-Y_{t,r,\theta,c})^\beta (\hat{Y}_{t,r,\theta,c})^\alpha \log (1-\hat{Y}_{t,r,\theta,c}) & \text{if} \; Y_{t,r,\theta,c} = 0  \;\\
            &\text{and} \; Y_{t,r,\theta,\Bar{c}} > 0 \\
            (1-Y_{t,r,\theta,c})^\beta (\hat{Y}_{t,r,\theta,c})^\alpha \log (1-\hat{Y}_{t,r,\theta,c}) & \text{otherwise}
        \end{cases}
        \label{loss equ}
    \end{aligned}$}
\end{equation}

\noindent where $\alpha$ and $\beta$ are hyper-parameters of focal loss \cite{lin2017focal}, and $N_{\R{obj}}$ is the number of objects in ground truth. The normalization by $N_{\R{obj}}$ is chosen as to normalize all positive focal loss instances to 1. Compared to \cite{lin2017focal}, we add a new scalar hyper-parameter $\kappa$, which put more loss/focus at the region where objects exist to shorten the training time. In this paper, we choose $\kappa = 4$ and following \cite{zhou2019objects}, we use $\alpha = 2$ and $\beta = 4$ in all our experiments.

When designing the loss function, we also take account of the fact that NN may well give more weights to the straightforward and accessible RA features than other velocity-based features, leading to overfitting. This point can be illustrated with the above human perception example again. A person with unimpaired eyesight will not rely much on the other sensors (ears, nose), thus resulting in weaker supplementary function compared to a person with impaired eyesight \footnote{To train this supplementary function for a non-disabled person, it is better to create a situation where eyes are not working, e.g., blindfolding}. 

The above analogy applied to loss function design suggests how to make NN fully utilize all three perspectives (RA, RV and VA) and particularly enhance the supplementary function provided by RV and VA perspective. We add a new loss constraint $L_{\hat{Y}^\prime Y}$ besides the original loss term $L_{\hat{Y}Y}$ mentioned above. To obtain $L_{\hat{Y}^\prime Y}$, we set $X_{\R{RA}} = 0$ in the new loss term, i.e. we input $X=\left(0, X_{\R{RV}}, X_{\R{VA}}\right)$ to the NN $P_w(Y|X)$ again such that getting the new prediction $\hat{Y}^\prime$. Then $\hat{Y}^\prime$ is also supervised by ground truth $Y$ with \eqref{loss equ} to obtain $L_{\hat{Y}^\prime Y}$.

The final loss is the weighted sum of two terms:
\begin{equation}
L_{loss} = L_{\hat{Y}Y} + \gamma L_{\hat{Y}^\prime Y}
\end{equation}

\noindent where $\gamma$ is the hyper-parameter to balance two terms, chosen to be $\gamma=0.5$ in this paper.

\begin{figure}
    \centering
    \includegraphics[width=3.1in]{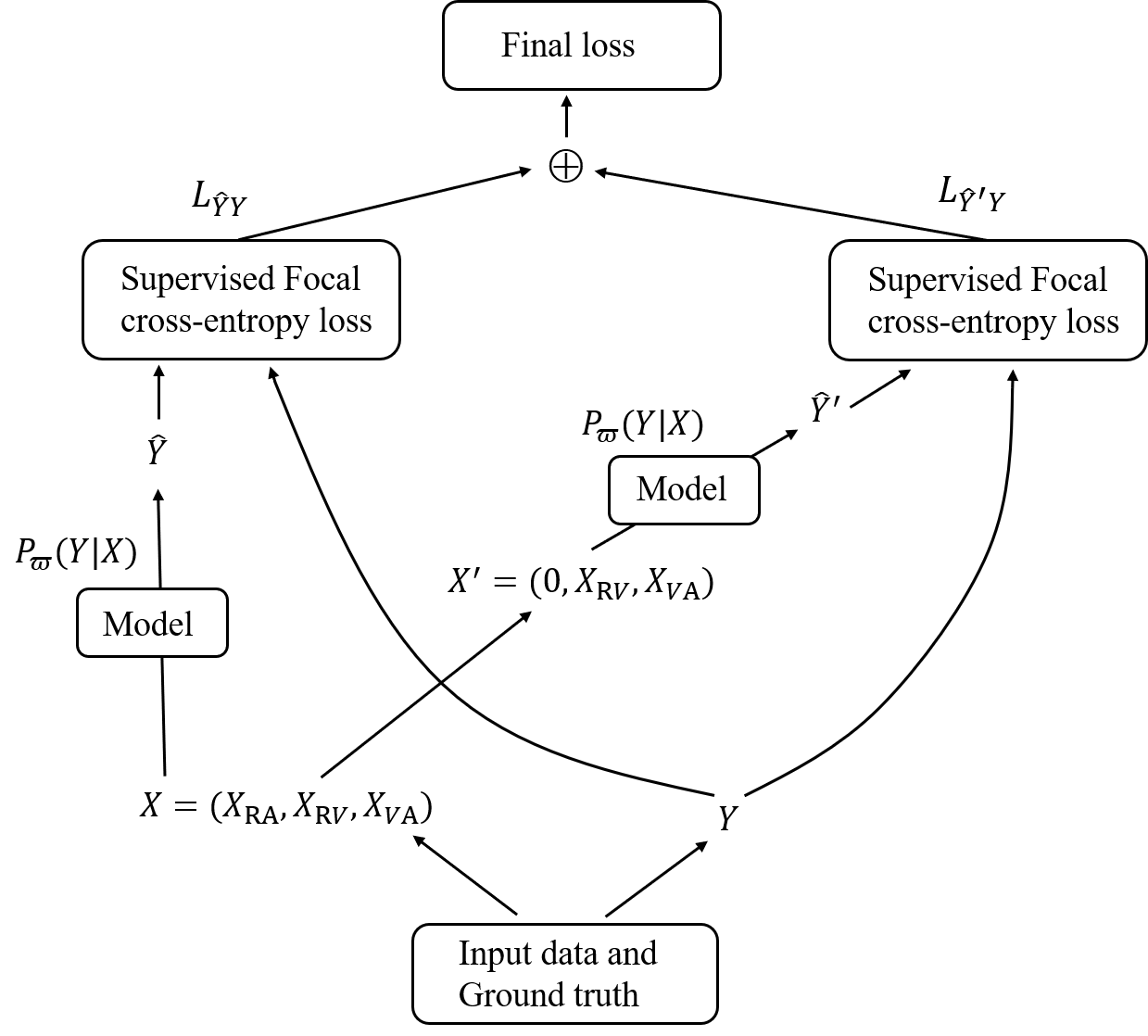}
    \caption{The loss function for all-perspectives supervision}
  \label{loss}
  \vspace{-3mm}
\end{figure}

\section{Radar Data Augmentation Algorithms \label{radar_aug}} 

Many image data augmentation algorithms have been proposed to increase the amount of {\em relevant} data and prevent the NN from overfitting, thus essentially boosting overall performance. However, most of the existing data augmentation algorithms cannot be applied to the radar data because of a few key differences from traditional RGB images - complex inputs, energy loss with range, and nonuniform resolution in the angular domain. In this section, we focus on 4 basic data augmentation operations and explain how to apply them to radar data: \highlighttext{flipping, translating, interpolating, and mixing}. 

\subsubsection*{\textbf{Horizontal Flipping}}
The horizontal flipping operation will swap the left and right part of the processed 3D radar cube along azimuth angle dimension. This operation can be applied to radar data directly as to images because radar has symmetric property (resolution and antenna gain) in the angular domain.

\begin{figure*}[h]
    \centering
    \resizebox{\textwidth}{!}{\includegraphics[width=6in,trim=1 1 1 1,clip]{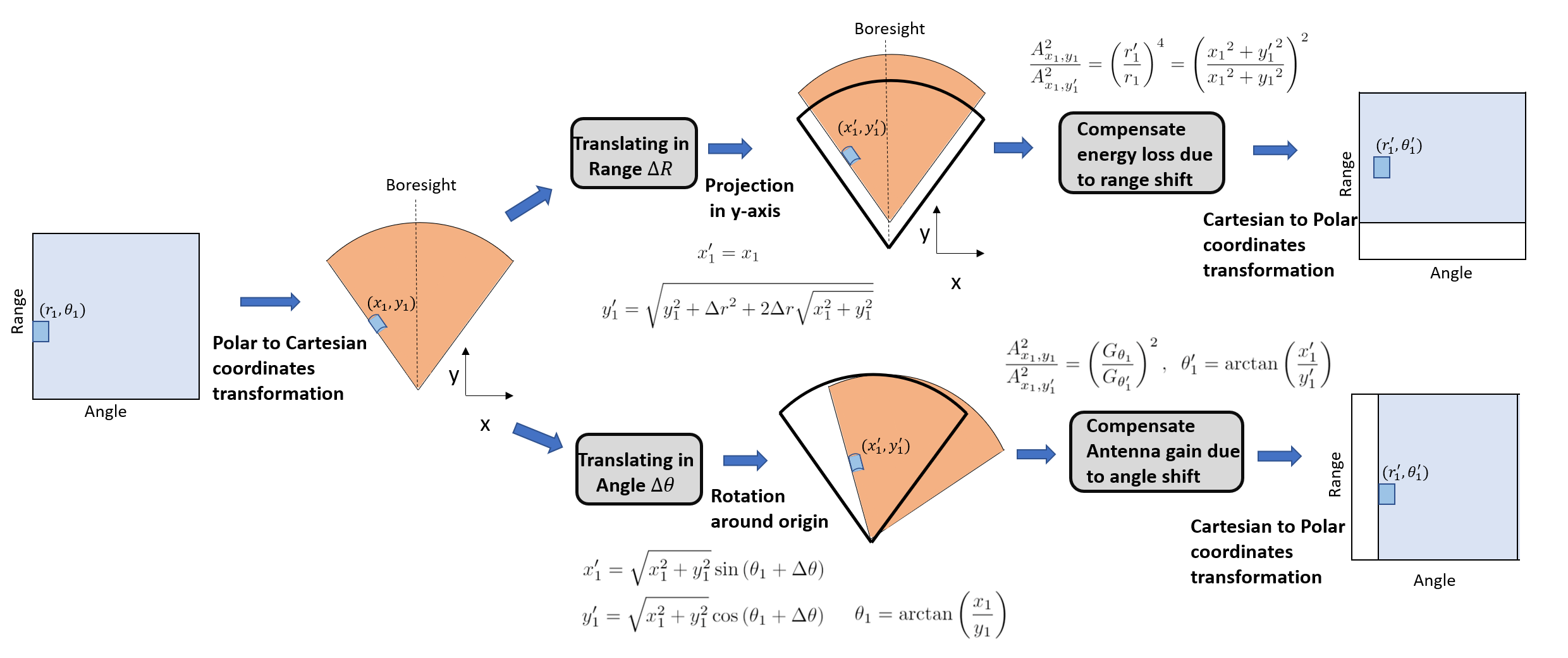}}
    \caption{The design of translating in range and angle}
    \label{trans}
    \vspace{-3mm}
\end{figure*}

\subsubsection*{\textbf{Translating in Range}}

With the translating in range operation, we will do a pre-defined range shift $\Delta r$ for all objects in RA domain. As shown in Fig. \ref{trans}, for the first step, we transform the polar-coordinates\footnote{\label{nonuniform}The processed 3D radar cube is represented as the polar-coordinates format in RA domain, which is non-uniform for the representation of objects.} $(r, \theta)$ radar data into the uniform Cartesian-coordinates $(x, y)$ radar data with the well-known projection: $x = r \sin \theta, \; y = r \cos \theta$. This relation is nonlinear, therefore the new cell in Cartesian coordinates is not rectangular, interpolation and down-sampling operations are needed to sample the Cartesian plane uniformly.

Assume that there is a target at location $(r_1, \theta_1)$ with corresponding Cartesian-coordinates $(x_1, y_1)$, and the distance from target to radar boresight is fixed even with range shift. Then, the range shift $\Delta r$ for this target is equivalent to shift y-axis while keeping the x-axis fixed in the Cartesian plane. This maps the previous cell $(x_1, y_1)$ to the new cell $(x'_1, y'_1)$ with relation: $x'_1 = x_1,\; y'_1 = \sqrt{y_1^2 + {\Delta r}^2 + 2 \Delta r r_1}$, where $r_1 = \sqrt{x_1^2 + y_1^2}$, and $r'_1 = \sqrt{{x^\prime_1}^2 + {y^\prime_1}^2} = r_1 + \Delta r$.

Without considering the Doppler phase change, the translating in range operations can be approached by shifting $\lfloor - \frac{2M_r S \Delta r}{c_{\R{0}} f_{\R{s}}} \rceil$ cells in the polar-coordinates range spectrum/profile, and then changing the phase across antennas (see Appendix \ref{apx_trans_rang}):
\begin{equation}
    \frac{\phi_{r'_1, q}}{\phi_{r_1, q}}= \frac{qd \sin \theta_1^\prime}{qd \sin \theta_1} = \frac{r_1}{r_1 + \Delta r}
\end{equation}

\noindent where $M_r$ is the number of points for Range FFT, $\phi_{r_1, q}$ is the original phase of $q^{th}$ Rx for the target at $r_1$, $\phi_{r'_1, q}$ is the phase of $q^{th}$ Rx after projecting the target to $r'_1$. The phase of Rx changes as the azimuth angle of target changes when applying the translating in range operation.

Meanwhile, the energy loss due to range shift needs to be compensated according to the radar range equation \cite{doi:10.1036/0071444742}:
\begin{equation}
\frac{A_{x^\prime_1, y^\prime_1}}{A_{x_1, y_1}}= \left(\frac{r_1}{r^\prime_1}\right)^2 = \left(\frac{r_1}{r_1 + \Delta r}\right)^2
\end{equation}

\noindent where $A$ is the signal amplitude. 
 
\subsubsection*{\textbf{Translating in Angle}}
With the translating in angle operation, we will do a pre-defined angle shift $\Delta \theta$ for all objects in RA domain.
The angle shift in polar plane is equivalent to the rotation around origin in Cartesian plane. Therefore, as shown in Fig. \ref{trans}, after transforming to the Cartesian-plane data, we use the following relation to map the target in cell $(x_1,y_1)$ to the new cell $(x^\prime_1, y^\prime_1)$: $x'_1 = r_1 \sin\left(\theta_1 + \Delta \theta \right), \; y'_1 = r_1 \cos\left(\theta_1 + \Delta \theta \right)$, where $\theta_1 = \arctan{(\frac{x_1}{y_1})}$, and $\theta^\prime_1 = \arctan{(\frac{x'_1}{y'_1})} = \theta_1 + \Delta \theta$.

If there is no more than one target in a range bin, we can approximate above translating in angle operation by shifting $\lfloor \frac{M_{\theta} d \left(\sin{\theta_1} - \sin{\theta^{\prime}_1} \right)}{\lambda} \rceil$ cells in the polar-coordinates angular spectrum (see Appendix \ref{apx_trans_ang}), where $M_\theta$ is the number of points for Angle FFT.

Based on radar range equation \cite{doi:10.1036/0071444742}, we also need to compensate for the antenna gain $(G)$ loss due to angle shift:
\begin{equation}
\frac{A_{x^\prime_1, y^\prime_1}}{A_{x_1, y_1}}=\frac{G_{\theta^\prime_1}}{G_{\theta_1}}
\end{equation}

\subsubsection*{\textbf{Interpolating}}
The interpolating operation is to fill in the blanks (the white stripes in the last two images of Fig. \ref{trans}) left by the translating operation. We utilize the environment noise for interpolating in order to imitate the situation where there is no object in the blank area. The environment noise samples are obtained by sorting all data of the 3D radar cube with amplitude and then taking the bottom (smallest) $5\%$ of it. 

\subsubsection*{\textbf{Mixing}}
The mixing operation is to add up two 3D radar cubes, which could remain unchanged or could have been done with other augmentation techniques - like flipping and translating.

\section{Experiment \label{experiments}}
\subsection{UW Camera-Radar (CR) Dataset}
\begin{figure}
    \centering
    \includegraphics[width=3.2in]{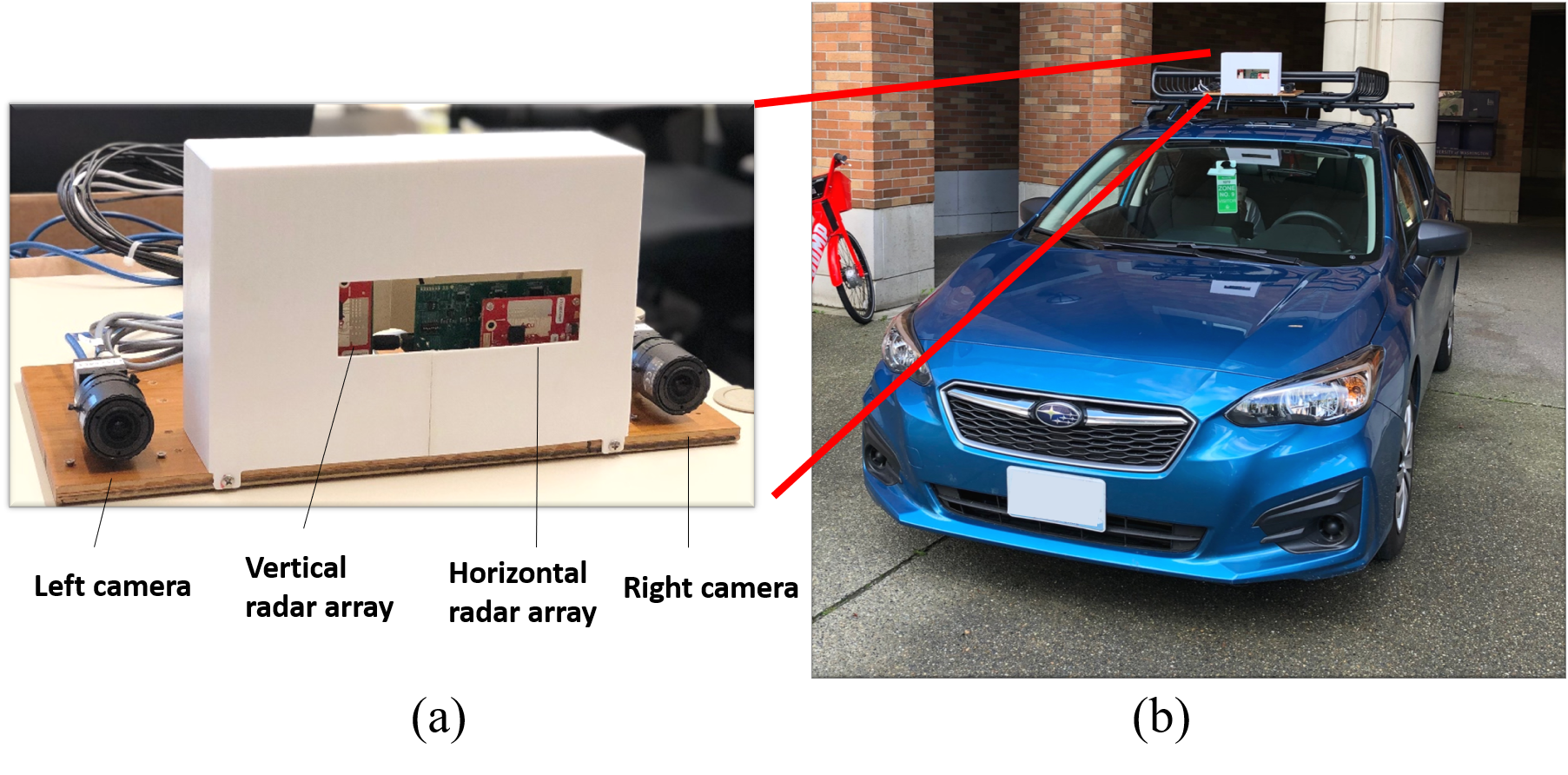}
    \caption{Radar-camera data capture platform: (a) This platform consists of 2 FLIR cameras and two perpendicular radars from TI - the right radar is with the 1D horizontal antenna array, and the left one is with the 1D vertical antenna array. (b) Data capture platform mounted on a vehicle with front view.}
   \label{platform}
    \vspace{-3mm}
\end{figure}

\begin{figure}
    \hspace{-0.5em}
    \includegraphics[width=3.4in]{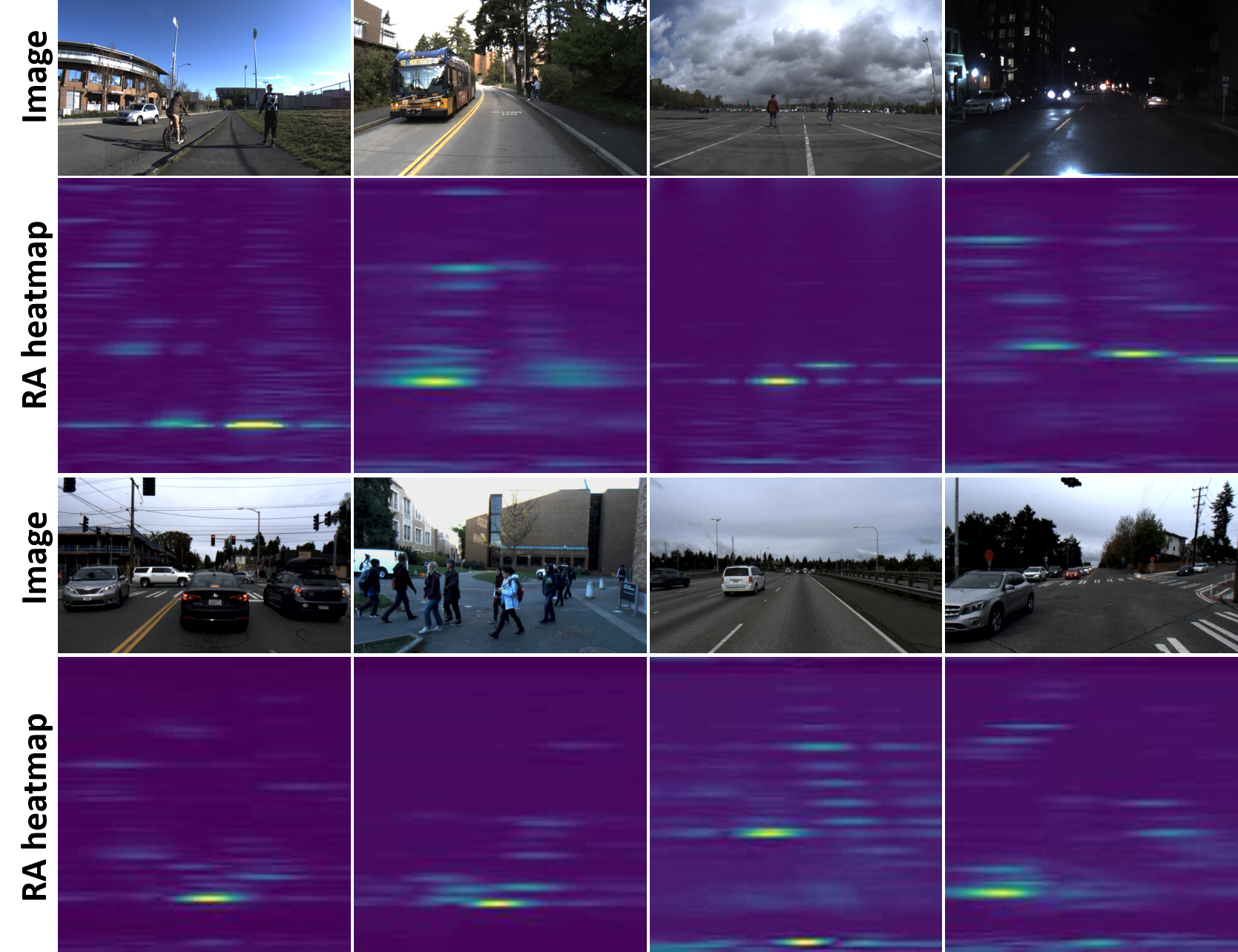}
    \caption{8 scenario examples in the collected UWCR dataset: row 1, 3 are the camera images; row 2, 4 are the corresponding radar range-azimuth angle heatmaps.}
    \label{dataset}
    \vspace{-3mm}
\end{figure}

A large camera image and radar raw data (I-Q samples post demodulated at the receiver) dataset for various objects have been collected for multiple scenarios -  parking lot, curbside, campus road, city road, freeway, etc. - by a vehicle-mounted platform that is driven (see Fig. \ref{platform}(b)). In particular, significant effort was placed in collecting data for situations where cameras are largely ineffective, i.e. under challenging light conditions \footnote{Please contact us if you are interested in the dataset. (xygao@uw.edu)}. We show the camera images and radar range-angle heatmaps of several scenario examples in our UWCR dataset at Fig. \ref{dataset}. 

The data collection platform shown in Fig. \ref{platform}(a) consists of 2 FLIR cameras (left and right) and two TI AWR1843 EVM radars \cite{ti1843evm}. Two radar EVM boards are placed to form a ‘2D’ antenna array system\footnote{Here, `2D' is equivalent to two perpendicular 1-D arrays.} that can provide more abundant object information. We place one radar array horizontally and the other one vertically to collect the data from both range-azimuth angle domain and range-elevation angle domain. We only use the radar data from horizontal array so far, and we will incorporate the vertical array data into the future work.

As discussed in Section \ref{cam sup}, the binocular cameras are synchronized with radars, and they can provide the location and class of semantic objects after we implement the Mask R-CNN detection model \cite{he2017mask} and unsupervised depth estimation model \cite{monodepth17, 10.1145/3343031.3350924} on the captured camera images. The semantic object detection results and depth estimation results generated from cameras are manually calibrated and then saved as the requisite ground truth for the following training and evaluation.

\begin{table}
  \caption{Dataset distribution for training and test}
  \label{data train test}
  \centering
    \resizebox{0.5 \textwidth}{!}{
    \begin{tabular}{cccc}
    \toprule
    & Augmented data & Training set \footnote{The training set here doesn't count the augmented data.} & Testing set\\
    \midrule
    Frames  & 26462 & 67198 & 25098\\
    \midrule
    Included ped. cyc. car \footnote{Pedestrian (ped), cyclist (cyc).} & 53275, 18840, 18731 &  55203, 24742, 46446 & 14541, 7607, 7471\\
  \end{tabular}}
  
  \resizebox{0.5 \textwidth}{!}{
    \begin{tabular}{ccccc}
    \toprule
    & Parking Lot & Curbside & On-road & nighttime\\
    \midrule
    Frames  & 9900 & 7200 & 4398 & 3600\\
    \midrule
    Included ped. cyc. car  & 5750, 4501, 2700 & 3581, 1172, 2039 & 5210, 1934, 2732 & - \footnote{The nighttime data are not labeled, so we don't count the number of different classes of objects here.}\\
    \bottomrule
  \end{tabular}}
  \vspace{-3mm}
\end{table}

\subsection{Data Processing}
\subsubsection*{3-DFFT Preprocessing}
We implement the 3-DFFT \cite{gao2019experiments} algorithm on the raw I-Q radar data samples to obtain the RVA heatmap sequences. The FFT on range, angle, velocity dimensions are all 128 points. We choose input frame number $M=16$. Therefore, the size of the preprocessed input data is $128\times 128 \times 128  \times 16$ that corresponds to $[\text{range bins}\times \text{angle bins} \times \text{velocity bins} \times \text{frame number}]$.

\subsubsection*{Data Augmentation}

We implement the proposed data augmentation algorithms on the processed RVA heatmap sequences, which include flipping, range translating, angle translating, and mixing the input data. The augmented data is saved locally and \textbf{only} used as part of the training data to avoid overfitting.

\subsubsection*{Training and Test sets}

We partition our UWCR dataset into the training set and test set. Any nighttime scenario data cannot be used in the training set as the corresponding low-light camera images cannot provide the ground truth labels. So all nighttime data is placed into the test set for qualitative performance evaluation only, i.e, the performance of nighttime data isn't evaluated with numerical metrics. 

The data distribution for the training, augmented, and test set are shown in Table. \ref{data train test} (row 1-3). Note that the training set in table doesn't count augmented data, and the whole training data are the collection of training set and augmented data. The test set is divided into 4 scenarios: parking lot, curbside, on-road, and nighttime. Table. \ref{data train test} also shows their data distribution.

\begin{figure*}[h]
    \hspace{-1.5em}
    \includegraphics[width=7.5in]{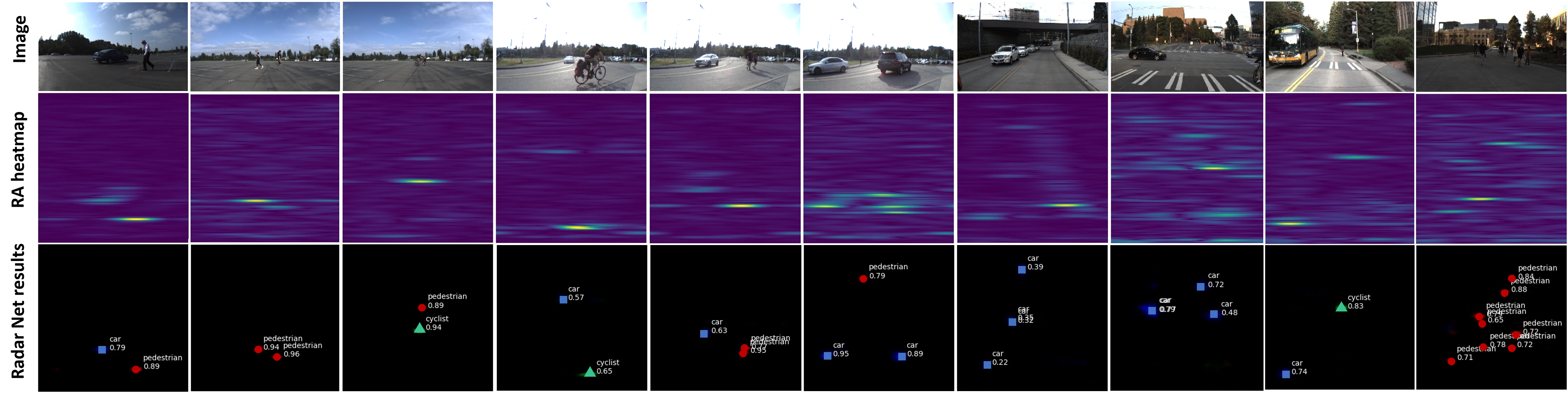}
    \caption{10 test examples from the parking lot scenario, curbside scenario, and on-road scenario: Column 1-3 are the parking lot scenario; Column 4-6 are the curbside scenario; Column 7-10 are the on-road scenario. For each column, the top row image is the synchronized camera image for visualization, the second row image is the corresponding radar RA heatmap, and the bottom row image is the visualization of the RAMP-CNN model results.}
  \label{res}
    \vspace{-3mm}
\end{figure*}

\begin{table*}[h]
  \caption{Performance comparison between different models}
  \label{perfor}
  \centering
    \begin{tabular}{ccccccccc}
    \toprule
    Model   &\multicolumn{2}{c}{Overall}   &\multicolumn{2}{c}{Parking Lot Scenario}   &\multicolumn{2}{c}{Curbside Scenario}    &\multicolumn{2}{c}{On-road Scenario}\\
    \cmidrule(r){2-3} \cmidrule(r){4-5} \cmidrule(r){6-7}
    \cmidrule(r){8-9}
    & AP & AR & AP & AR & AP & AR & AP & AR \\
    \midrule
    CDMC \cite{gao2019experiments} & 30.55\%  & 54.79\%  & 65.74\% & 76.56\% & 28.68\% & 53.93\% & 4.88\% & 25.76\%\\
    \midrule
    RODNet-HG \cite{wang2020rodnet} & 71.84\%  &  76.03\% & 93.87\% & 95.36\% & 61.65\% & 70.09\% & 41.97\% & 53.04\%\\
    \midrule
    RODNet-CDC \cite{wang2020rodnet} & 71.46\%  & 78.15\%  & 92.72\% & 95.07\% & 64.01\% & 71.97\% & 46.52\% & 58.61\%\\
    \midrule
    Prop. RAMP-CNN & \textbf{81.23\%}  & \textbf{84.25\%}  & \textbf{97.38\%} & \textbf{98.37\%} & \textbf{79.25\%} & \textbf{84.21\%} & \textbf{57.07\%} &
    \textbf{64.85\%}\\
    \bottomrule
  \end{tabular}
\end{table*}

\subsection{Experiments \label{train detail}}
\subsubsection*{\textbf{Baselines}} 
We compare the RAMP-CNN model with RODNet-CDC \cite{wang2020rodnet}, RODNet-HG \cite{wang2020rodnet}, the state-of-art radar object classification models, as well as the CDMC \cite{gao2019experiments}, a model that fully exploits the micro-Doppler signatures of moving objects.
\subsubsection*{\textbf{Training}}
We train the RAMP-CNN model and retrain the RODNet-CDC, RODNet-HG, CDMC following below details. 

\subsubsection{Proposed RAMP-CNN model}
We train the RAMP-CNN model on complete training set (includes augmented data) with Cyclic learning rate (minimum $5 \times 10^{-6}$, maximum $5\times 10^{-5}$, and cycle duration 860 iterations) \cite{7926641}, batch size 5 for the first 10 epochs. Then we continue to train the RAMP-CNN model with Step learning rate (starts from $5 \times 10^{-6}$, and decays 0.2 every 5 epochs), batch size 4 for the next 24 epochs. We use the Adam gradient descent optimizer \cite{adam} and 1 TITAN RTX GPU for the training of all experiments. To verify the capability of the proposed radar data augmentation algorithms (see Section \ref{ablation study}), we also train a new RAMP-CNN model following the same procedures as above, but without augmented data.  

\subsubsection{RODNet-CDC and RODNet-HG}
We train the RODNet-CDC and RODNet-HG model with Cyclic learning rate (same as above), batch size 4 for 10 epochs, and then train them with Step learning rate (same as above), batch size 4 for the following 22 epochs. The gradient descent optimizer is Adam \cite{adam}. The loss function for RODNet-CDC is the Minimum Square Error (MSE) provided by PyTorch, and the loss function for RODNet-HG is Cross Entropy. Note that the training set for RODNet-CDC, RODNet-HG and CDMC \cite{gao2019experiments} model doesn't include the augmented data.

\subsubsection{CDMC}
Following \cite{gao2019experiments}, we generated about $1.2\times10^5$ concatenated STFT heatmaps in total from the training set. By feeding the training STFT heatmaps to the VGG16 classifier, we trained the model from scratch with the batch size 5, learning rate $1 \times 10^{-4}$ for the first 10 epochs,  and learning rate $1 \times 10^{-5}$ for the next 10 epochs.  The gradient descent optimizer is also Adam \cite{adam} and the loss function is the Cross Entropy provided by TensorFlow.


\subsection{Evaluation Metrics}
We use the average precision (AP) and average recall (AR) to evaluate performance - using true positive, false positive, and false negative rates in \eqref{defination}. Here, true positive ($\R{tp}$) represents correctly located and classified instances, false positive ($\R{fp}$) represents the false alarm, false negative ($\R{fn}$) represents the missed detection and/or incorrectly classified instance. 
\begin{equation}
    \text{Precision} = \frac{\R{tp}}{\R{tp}+\R{fp}}, \quad \text{Recall} = \frac{\R{tp}}{\R{tp}+\R{fn}}
    \label{defination}
\end{equation}

We adopt the CFAR \cite{doi:10.1036/0071444742} and threshold 0.2 to filter out the target center points from prediction $\hat{Y}$. Whether the targets are correctly located is determined by a object size-adaptive distance threshold, i.e., if the distance between the prediction and ground truth is smaller than the threshold, we assume the prediction is correctly located.

\subsection{Evaluation Results \label{test set sec}}
We test the RAMP-CNN under 4 different scenarios: parking lot, curbside, on-road, and nighttime (See Fig. \ref{res} and Fig. \ref{res2} for testing examples). The parking lot scenario is manually controlled to have moving and/or static objects at a clear parking lot. For the curbside scenario, we set up the data collection platform on the curbside and then record multiple moving objects on a clear road. The on-road scenario is more like an autonomous driving scenario where we drive around and record all objects on the city road. For the nighttime scenario, we collect the data under challenging light conditions where cameras are largely ineffective. The testing results of RAMP-CNN and other baselines are shown in Table. \ref{perfor}.
\subsubsection*{The Parking Lot Scenario}
The parking lot test set has the data of 9900 frames which contain 5750 pedestrians, 4501 cyclists, and 2700 cars. The parking lot scenario is relatively easy for the object recognition task as the background is clean and the objects are few. The RAMP-CNN model achieves \textbf{nearly perfect} performance ($97.38\%$ AP, $98.37\%$ AR), and beat all prior works. We show 3 test examples in Fig. \ref{res}.

\subsubsection*{The Curbside Scenario}
The curbside test set has the data of 7200 frames which contain 3581 pedestrians, 1172 cyclists, and 2039 cars. This scenario allows multiple objects to appear at the same time and some of them to be close, so that it is harder than parking lot scenario. The AP ($79.25\%$) and AR ($84.21\%$) of RAMP-CNN model have \textbf{around $\textbf{15}\%$ improvement} over the best results of prior work - RODNet-CDC model ($64.01\%$ AP, $71.97\%$ AR). We show 3 test examples in Fig. \ref{res}.

\subsubsection*{The On-road Scenario}
The on-road test set has the data of 4398 frames which contain 5210 pedestrians, 1934 cyclists, and 2732 cars. This test set is collected from the city-road driving experiments which include several challenging situations, e.g., the strong reflections from the environment, a large number of cars in the field of view, crowded pedestrians, etc. The RAMP-CNN model obtains $\textbf{10}\%$ \textbf{improvement in AP} and $\textbf{6}\%$ \textbf{improvement in AR} over the RODNet-CDC baseline. We show 3 test examples in Fig. \ref{res}.

\subsubsection*{The Nighttime Scenario}
\begin{figure}
    \hspace{-0.5em}
    \includegraphics[width=3.3in]{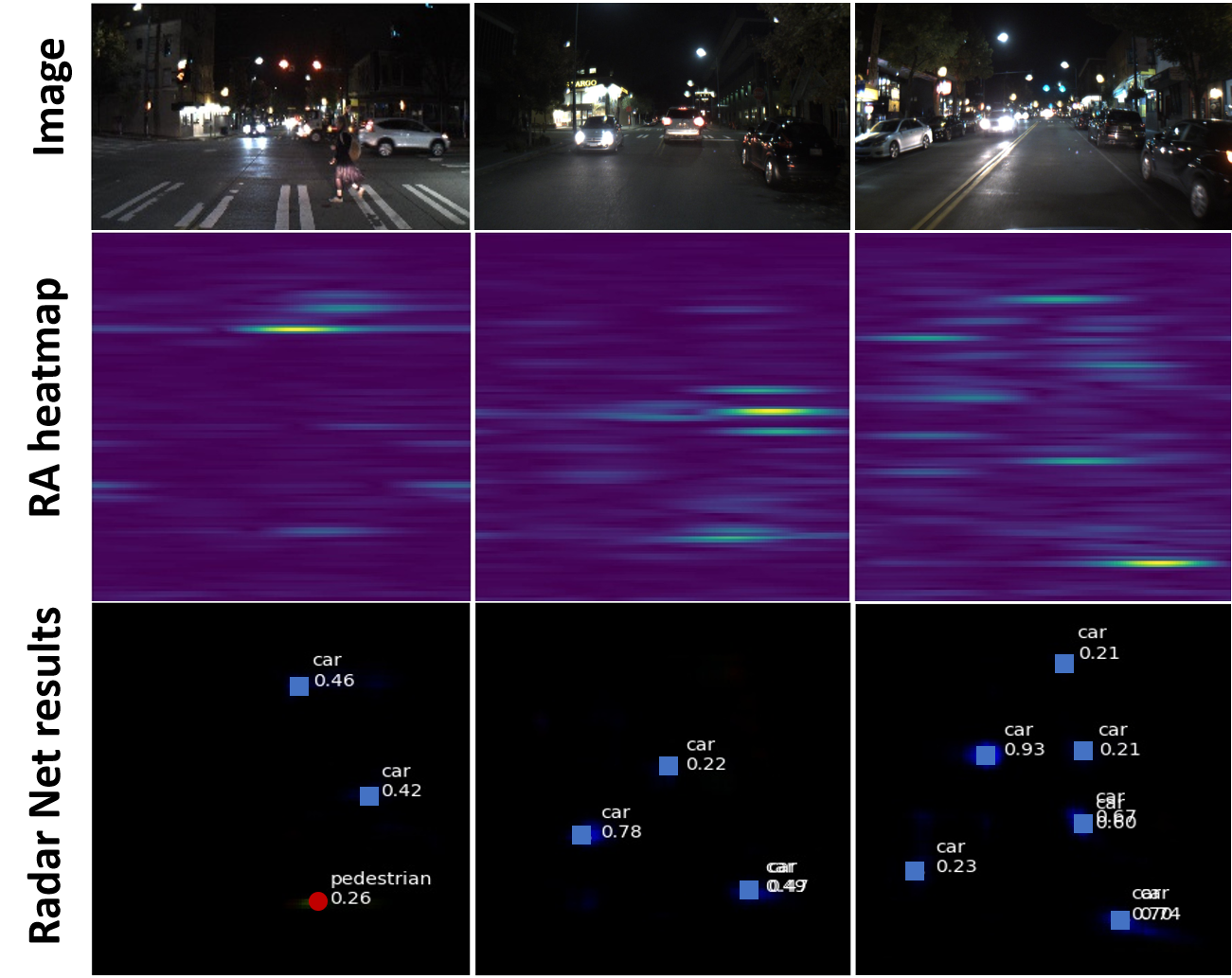}
    \caption{3 test examples from the nighttime scenario. The arrangement is same as Fig. \ref{res}.}
  \label{res2}
    \vspace{-3mm}
\end{figure}

We test the RAMP-CNN under nighttime to support the objective of this paper - advance the cause of radar as a low-cost substitute for optical sensors that fail under such severe conditions. As shown in Fig. \ref{res2}, the RAMP-CNN model performs as well as under the daytime scenario, i.e. radar is impervious/robust to sunlight change. As it is hard to implement the ground truth labeling on nighttime set, we don't numerically evaluate the performance here.

\section{Analysis and Ablation study \label{Analysis}}

\subsection{Impact of adding Temporal information}
Compared to prior works, the proposed RAMP-CNN model fully exploits the temporal information behind the chirps within one frame, as well as the change of spatial information (range-angle info.) across frames; hence we expect it to essentially achieve performance improvements for moving objects. To verify this, we choose a part of the data from the parking lot and curbside scenario, and redivide them into the static object set and moving object set. The distribution of these two sets is shown in Table. \ref{static mov dist}. 

We evaluate the performance of RAMP-CNN, CDMC, RODNet-HG, and RODNet-CDC model on the static object set and moving object set respectively. From the evaluation results (Table. \ref{perfor_staticmov}), we know that for static object scenario, the performance of RAMP-CNN (AP around $67\%$, AR around $70\%$) in the same level with RODNet-CDC, but much better than RODNet-HG model; for moving object scenario, there is a performance gap (about $15\%$ AP and $13\%$ AR) between RAMP-CNN model (AP $80\%$, AR $84\%$) and other baselines (AP around $65\%$, AR around $71\%$). The results above verify that the added temporal information in RAMP-CNN model is helpful for the recognition with moving objects.

\begin{table}[ht]
  \vspace{-3mm}
  \caption{Dataset distribution for static object scenario and moving object scenario}
  \label{static mov dist}
  \centering
  \resizebox{0.48\textwidth}{!}{
    \begin{tabular}{ccc}
    \toprule
     & Static Object Scenario & Moving Object Scenario\\
    \midrule
    \highlighttext{Frames}  & 3600 & 7200 \\
    \midrule
    \highlighttext{Included ped. cyc. car} & 1250, 1800, 1800 & 2883, 1083, 1702 \\
    \bottomrule
  \end{tabular}}
 \vspace{-3mm}
\end{table}

\begin{table}[ht]
   \vspace{-3mm}
  \caption{The performance of RAMP-CNN model for static object scenario and moving object scenario}
  \label{perfor_staticmov}
  \centering
    \begin{tabular}{ccccc}
    \toprule
    Model   &\multicolumn{2}{c}{Static Object Scenario}   &\multicolumn{2}{c}{Moving Object Scenario}\\
    \cmidrule(r){2-3} \cmidrule(r){4-5} & AP & AR & AP & AR\\
    \midrule
    CDMC \cite{gao2019experiments} & 32.34\% & 50.73\% & 28.68\% & 53.93\% \\
    \midrule
    RODNet-HG \cite{wang2020rodnet} & 56.26\% & 61.36\% & 61.68\% & 70.11\%\\
    \midrule
    RODNet-CDC \cite{wang2020rodnet} & 65.02\% & 69.91\% & 64.01\% & 71.97\%\\
    \midrule
    RAMP-CNN & \textbf{67.58\%} & \textbf{70.27\%} & \textbf{79.25\%} & \textbf{84.21\%}\\
    \bottomrule
  \end{tabular}
  \vspace{-3mm}
\end{table}

\begin{table*}
  \vspace{-3mm}
  \caption{Ablation Study}
  \label{ablation}
  \centering
    \begin{tabular}{ccccccccccc}
    \toprule
    Model   & Data Augmentation  
    & New training loss &\multicolumn{2}{c}{Overall} &\multicolumn{2}{c}{Parking lot scenario}   &\multicolumn{2}{c}{Curbside scenario}    &\multicolumn{2}{c}{On-road scenario}\\
     \cmidrule(r){4-5} 
     \cmidrule(r){6-7} \cmidrule(r){8-9}
     \cmidrule(r){10-11}
     & & & AP & AR & AP & AR & AP & AR & AP & AR\\
    \midrule
    RAMP-CNN &   & \Checkmark & 76.78\% & 81.39\% & 95.91\% & 97.21\% & 66.47\% & 75.83\% & 52.79\% &
    62.66\%\\
    \midrule
    RAMP-CNN & \Checkmark  &  & 76.93\% & 81.41\% & 93.90\% & 95.46\% & 75.52\% & 81.81\% & 54.10\% &
    61.86\%\\
    \midrule
    RAMP-CNN &  \Checkmark & \Checkmark &
    \textbf{81.23\%} & \textbf{84.25\%} & \textbf{97.38\%} & \textbf{98.37\%} & \textbf{79.25\%} & \textbf{84.21\%} & \textbf{57.07\%} &
    \textbf{64.85\%}\\
    \bottomrule
  \end{tabular}
  \vspace{-3mm}
\end{table*}

\subsection{Ablation Study \label{ablation study}}
An ablation study refers to removing some “part/module” of the model or algorithm, and seeing how that affects performance. In this section, we will study the contribution of two parts of RAMP-CNN model - data augmentation and new training loss - to the performance. 

Following the training procedure mentioned in Section \ref{train detail}, we train two `incomplete' RAMP-CNN models - one removes the data augmentation part, the other one replaces the proposed training loss with the ordinary focal loss \cite{lin2017focal, zhou2019objects}. The trained models are all evaluated on 4 test sets mentioned above for comparison with the performance of the 'complete' RAMP-CNN model (presented in Table. \ref{perfor}). The experiment results of the ablation study are shown in Table. \ref{ablation}. 

The experiments between the RAMP-CNN model with and without data augmentation confirm that the proposed augmentation algorithms help to boost model performance by avoiding overfitting. For illustration, the data augmentation algorithms make RAMP-CNN get $12\%$ AP improvement and $8\%$ AR improvement in the curbside scenario, and get $4\%$ AP improvement and $2\%$ AR improvement in the on-road scenario.

The experiments between the RAMP-CNN model with and without the proposed training loss function verify that the proposed training loss helps improve performance by pushing the RAMP-CNN model to learn more Doppler-related features. Specifically, RAMP-CNN obtains around $4\%$ AP improvement and $3\%$ AR improvement in both parking lot scenario and curbside scenario as well as the on-road scenario. 

\subsection{Complexity Analysis}
In this section, we analyze the time complexity and space complexity of different models. 

\subsubsection*{\textbf{Time complexity}}
Time complexity is the amount of time it takes to run the algorithm. We count the number of floating-points operations (\textbf{FLOPs}) required by algorithm, to measure time complexity. The time complexity of the overall CNN is the sum of the time complexity of all {\em conv} layers \footnote{The time cost of fully connected layers and pooling layers is not involved in the this formulation. These layers typically take
5-10\% computational time.} \cite{he2014convolutional}:
\begin{equation}
    \textbf{Time} \sim \mathcal{O}\left( \sum_{l=1}^{N_{\R{conv}}} I_l^n \cdot K_l^n \cdot C_{l-1} \cdot C_{l} \right)
\end{equation}

\noindent where $N_{\R{conv}}$ is the number of {\em conv} layers, $n$ is the dimension of convolution kernels (1-dim, 2-dim or 3-dim convolution), $I$ is the size of feature map, $K$ is the size of convolution kernel, $C_l$ is the number of output channels of the $l^{th}$ {\em conv} layer, that is, number of convolution kernels of this layer.

Another indicator to measure a model's time complexity is the training or prediction time. If time complexity is too high, it will lead to a large amount of time for model training and prediction. Therefore, we also measure the frame-level prediction/testing time for different models to evaluate the time complexity. We show the results in Table. \ref{time complexity}.

\subsubsection*{\textbf{Space complexity}}
Space complexity quantifies the amount of memory needed by an algorithm to run as a function. This consists of two parts: the total number of parameters (first term of \eqref{cal space comp}), and the occupied memory of the feature map output at all layers (second term of \eqref{cal space comp}).
\begin{equation}
    \label{cal space comp}
    \textbf{Space} \sim \mathcal{O}\left( \sum_{l=1}^{N_{\R{conv}}} K_l^n \cdot C_{l-1} \cdot C_{l} + \sum_{l=1}^{N_{\R{conv}}} I_l^n \cdot C_{l}\right)
\end{equation}

We show the space complexity results, as well as the number of layers for different models in Table. \ref{space complexity}

From Table. \ref{time complexity} and Table. \ref{space complexity}, we know that compared to the 4D-CDC model, RAMP-CNN needs almost \textbf{100} times fewer FLOPs, around \textbf{half} amount of  parameters, and \textbf{35} times smaller feature map size. For practical application, this means RAMP-CNN would not only run 100 times faster than 4D-CDC for both training and prediction, but also take 35 times less memory. That confirms the claimed statement - RAMP-CNN has much less computation complexity than the 4D model.

Also, compared to RODNet-CDC model, the time and space complexity of RAMP-CNN is around 3 times higher. That, however, means the performance improvement of RAMP-CNN model comes at the expense of increased complexity.

\begin{table}[h]
    \vspace{-3mm}
    \caption{Time Complexity Analysis}
    \label{time complexity}
    \centering
    \begin{tabular}{ccc}
    \toprule
    Model & FLOPs & Prediction time (per frame)\\
    \midrule
    RODNet-CDC &  $4.75\times 10^{11}$ & 11.2 ms\\
    \midrule
    4D-CDC \footnote{To compare the complexity between one 4D model and RAMP-CNN model model, we replace the 3D convolution kernels in RODNet-CDC model with the 4D convolution kernels and call the new model 4D-CDC. \label{4dcdc}} &  $ 1.64 \times 10^{14}$ & - \footnote{Note: we didn't implement the 4D-CDC model, so the prediction time and layer numbers are ignored here. \label{4D}}\\
    \midrule
    RAMP-CNN & $1.41\times 10^{12}$ & 31.1 ms\\
    \bottomrule
  \end{tabular}
  \vspace{-3mm}
\end{table}

\begin{table}[h]
  \caption{Space Complexity  Analysis}
  \label{space complexity}
  \centering
  \resizebox{0.48 \textwidth}{!}{
  \begin{tabular}{cccc}
    \toprule
    Model & Parameters amount & Feature map size & Layers number \footnote{The number of {\em conv} layers and {\em transposed conv} layers in models. For RODNet-CDC and RAMP-CNN model, the layers are all 3D; while for 4D-CDC model, all layers are 4D.}\\
    \midrule
    RODNet-CDC & $3.47\times 10^7$ & $6.31 \times 10^7$ & 6, 3\\
    \midrule
    4D-CDC  & $1.79\times 10^8$ & $6.58\times 10^9$ & 6, 3\\
    \midrule
    RAMP-CNN & $1.04\times 10^8$ &  $1.89 \times 10^8$ & 20, 9\\
    \bottomrule
  \end{tabular}}
  \vspace{-3mm}
\end{table}

\subsection{Summary}
The proposed RAMP-CNN model achieves significant performance improvement over prior works on object recognition under parking lot, curbside, and on-road scenario, \highlighttext{which establishes a new state-of-art baseline. In some hard cases, the radar object recognition functionality of RAMP-CNN might still be poor for supporting autonomous driving presently\footnote{\highlighttext{Based on the performance of camera object detection in \cite{suryAutodriv}, we infer that the acceptable/desired average precision and recall for radar would be 0.8 with different classes of objects.}}. However, it can be further improved in the future via incorporating more preprocessing to increase spatial resolution or adopting advanced radar platform with more antennas.}

RAMP-CNN is also verified to work at the nighttime scenarios, where cameras are largely ineffective due to the low-light. \highlighttext{Further, prior works \cite{YONEDA2019253, impactweather, fogexp} have been showing that mmW radars are with excellent environmental resistance and robustness because the millimeter-wave is less attenuated by fog, rain, or snow. Therefore, we have reason to believe that RAMP-CNN can be applied to these adverse conditions as a good substitute for optical sensors. However, due to the difficulty of capturing data in such circumstances locally, this must be left for future work.}

There are several other advantages of applying RAMP-CNN to mmW radars - it has excellent range localization ability because of the centimeter-level range resolution ($\sim$3.75 cm with 4 GHz sweep bandwidth). As shown in Fig. \ref{res} (column 10), RAMP-CNN can resolve multiple close pedestrians with range and localize them separately. Besides, RAMP-CNN model has great generalization for the input data with a higher dimension. For example, if we add the elevation dimension (from the vertical radar array) to the current RAMP-CNN input, then the formed 5D data can still be sliced and processed by several lower-dimension (3D) models that nonetheless achieve better performance with acceptable computation complexity.

RAMP-CNN model fully exploits the temporal information behind the chirps in one frame, as well as the change of spatial information (range-angle info.) across frames. Thus, the performance of RAMP-CNN particularly for moving objects, shows significant improvements relative to state-of-art.

The ablation study shows that both the proposed data augmentation algorithms and training loss are helpful for boosting the performance of RAMP-CNN. It is worth noting that major performance improvement comes from the main body of RAMP-CNN model (3-Perspectives model in Section \ref{RAMP-CNN model architecture}); the cumulative impact of all elements in the RAMP-CNN architecture results in the promising performance improvement, at the expense of increased complexity.

\section{Conclusion and Future Work \label{conclusion}}
In this paper, we propose a novel RAMP-CNN model for radar object recognition that can obtain the location (range and azimuth angle) and class of the objects in each frame by inputting the 3D radar cube sequences. The RAMP-CNN model fully exploits the temporal information behind the chirps in one frame, as well as the change of spatial information across frames, which makes it achieve significant performance improvement over the previous work. For future work, we are continuing to explore how to effectively utilize the radar data and create more sensible radar networks based on radar data properties.

\section*{Acknowledgement}
This work was supported by the CMMB Vision–UW EE Center on Satellite Multimedia and Connected Vehicles. The authors thank Yizhou Wang of Information Processing Lab (University of Washington) for the source codes of RODNet \cite{wang2020rodnet} that was used for the baseline evaluation in this paper.

\bibliography{bibtex}
\bibliographystyle{ieeetr}

\newpage
\appendix

\subsection{Translating in Range \label{apx_trans_rang}}

Suppose there is a target located at certain range $r_1$, azimuth angle $\theta_1$, the corresponding FMCW de-chirped signal (single chirp) at time $t$ and receiver $q$ can be expressed as \cite{Brooker_understandingmillimetre}:
\begin{equation}
    \resizebox{0.435\textwidth}{!}{
        $f(t,q) = A_1 \exp \left(j2\pi \left(f_{\R{c}}\tau_1 + S t \tau_1 - \frac{1}{2}S \tau_1^2 + \frac{q d\sin{\theta_1}}{\lambda} \right) \right)$
    }
    \label{dechirp}
\end{equation}

\noindent where $\tau_1=\frac{2r_1}{c_{\R{0}}}$ is the round-trip delay between radar and target, $A_1$ is the signal amplitude determined by target's radar cross section (RCS), range, and antenna gain.

With the Quadrature analog-to-digital conversion (ADC) sampling, the de-chirped signal can be expressed as $f(\frac{i}{f_{\R{s}}}, q)$, i.e., $t$ is replaced by the sampling time $\frac{i}{f_{\R{s}}}$, where $i$ is the sample index, and $f_{\R{s}}$ is the sampling frequency.

Implementing the Range FFT on the digitized de-chirped signal, we obtain the range profile $F_\R{R}(m_r, q)$ below:
\begin{equation}
\resizebox{0.435\textwidth}{!}{
$ \begin{aligned}
    F_\R{R}(m_{r}, q) = & \sum\limits_{i=0}^{N_{\R{s}}-1}
    f(\frac{i}{f_{\R{s}}}, q) \exp(-j\frac{2\pi m_{r} i}{M_r}) \\
    = & A_1 \exp \left(j2\pi \left(f_{\R{c}}\tau_1 - \frac{1}{2}S \tau_1^2 \right) \right) \exp \left(j2\pi \frac{q d\sin{\theta_1}}{\lambda} \right) \\
    & \times \left[\sum\limits_{i=0}^{N_{\R{s}}-1}   \exp \left(j2\pi S \frac{i}{f_{\R{s}}} \tau_1 \right) \exp(-j\frac{2\pi m_{r} i}{M_r}) \right] \\
    = & A_1 \exp \left(j2\pi \left(f_{\R{c}}\tau_1 - \frac{1}{2}S \tau_1^2 \right) \right) \exp \left(j2\pi \frac{q d\sin{\theta_1}}{\lambda} \right) \\
    & \times \frac{1 - \exp \left( j2\pi N_{\R{s}}\left( \frac{S \tau_1}{f_{\R{s}}} - \frac{m_{r}}{M_r} \right)\right)}{1-\exp \left( j2\pi \left( \frac{S \tau_1}{f_{\R{s}}} - \frac{m_{r}}{M_r} \right)\right)}\\
\end{aligned} $
}
\label{range_profile}
\end{equation}

\noindent where $M_r$ are the number of points for Range FFT, $m_r \in \{0, 1, ..., M_r-1\}$ is the range bin index. The mapping relationship between any range $r$ and bin index is shown below:
\begin{equation}
    m_r=\lfloor \frac{2M_r S r}{c_{\R{0}} f_{\R{s}}} \rfloor
    \label{rangedftmap}
\end{equation}

If the target at $(r_1, \theta_1)$ with corresponding Cartesian-coordinates $(x_1, y_1)$ has a translation in range, $\Delta r$, as we defined in Section \ref{radar_aug}, i.e. $x'_1 = x_1,\; y'_1 = \sqrt{y_1^2 + {\Delta r}^2 + 2 \Delta r r_1}$, then its new location $(r'_1, \theta^\prime_1)$ would be $r'_1 = r_1 + \Delta r$, $\theta^\prime_1 = \arctan(\frac{x'_1}{y'_1})$. According to radar range equation, the energy loss ratio due to range shift is $\left(r_1\right/\left(r_1 + \Delta r \right))^2$. Thus, the corresponding FMCW de-chirped signal $f_{\R{new}}(t,q)$ for the target at new location $(r'_1, \theta^\prime_1)$ is expressed as:
\begin{equation}
    \resizebox{0.435\textwidth}{!}{
    $f_{\R{new}}(t,q) = A_1 \left(\frac{r_1}{r_1 + \Delta r}\right)^2 \exp(j2\pi(f_{\R{c}}\tau'_1 + S t \tau'_1 - \frac{1}{2}S {\tau^\prime_1}^2 + \frac{qd\sin{\theta^\prime_1}}{\lambda}))$
    }
\end{equation}

\noindent where $\tau'_1=\frac{2r'_1}{c_{\R{0}}}$ is the new round-trip delay between radar and target.

The range profile for the digitized signal $f_{\R{new}}(\frac{i}{f_{\R{s}}},q)$ can be obtained with similar derivations in \eqref{range_profile} and \eqref{rangedftmap}. That is,
\begin{equation}
\resizebox{0.435\textwidth}{!}{
    $\begin{aligned}
    F_\R{R, new}(m_{r}, q) = & A_1 \left(\frac{r_1}{r_1 + \Delta r}\right)^2 \exp \left(j2\pi \left(f_{\R{c}}\tau'_1 - \frac{1}{2}S{\tau'_1}^2 \right) \right) \\
    & \times \exp \left(j2\pi \frac{q d\sin{\theta^\prime_1}}{\lambda} \right) \frac{1 - \exp \left( j2\pi N_{\R{s}}\left( \frac{S \tau'_1}{f_{\R{s}}} - \frac{m_{r}}{M_r} \right)\right)}{1-\exp \left( j2\pi \left( \frac{S \tau'_1}{f_{\R{s}}} - \frac{m_{r}}{M_r} \right)\right)}
    \end{aligned}$ 
}
\label{new_range_profile}
\end{equation}

Based on the triangular geometry, the $\sin{\theta'_1}$ term in \eqref{new_range_profile} can be simplified by $\sin{\theta'_1} = \frac{x'_1}{r'_1} = \frac{x_1}{r'_1}$. Ignoring the change of Doppler phase term $\exp \left(j2\pi \left(f_{\R{c}}\tau'_1 - \frac{1}{2}S {\tau'_1}^2 \right) \right)$, we would approximate the remaining terms of \eqref{new_range_profile} with the original profile \eqref{range_profile} by following relation:
\begin{equation}
\resizebox{0.435\textwidth}{!}{
    $\begin{aligned}
    F_{\R{R, new}}(m_{r}, q) \approx & \left(\frac{r_1}{r_1 + \Delta r}\right)^2 \left| F_{\R{R}}(m_{r}+\Delta m_{r_1}, q) \right| \\
    & \times \exp \left(j \frac{r_1}{r_1 + \Delta r} \phi \left(F_{\R{R}}(m_{r}+\Delta m_{r_1}, q) \right) \right)
    \end{aligned}$ 
}
\label{translate_range_equation}
\end{equation}

\noindent where $\Delta m_{r_1} = \lfloor -\frac{2M_r S \Delta r}{c_{\R{0}} f_{\R{s}}} \rceil$, $\phi \left(F_{\R{R}}(m_{r}+\Delta m_{r_1}, q) \right)$ is the phase of $F_{\R{R}}(m_{r}+\Delta m_{r_1}, q)$.

Equation \eqref{translate_range_equation} implies that we would obtain the new range spectrum $F_{\R{R, new}}(m_r, q)$ by shifting $\Delta m_{r_1}$ cells in the original range spectrum $F_{\R{R}}(m_r, q)$, and then scale up or down its amplitude and phase with coefficient $\left(\frac{r_1}{r_1 + \Delta r}\right)^2 $, $\frac{r_1}{r_1 + \Delta r}$ respectively. When there are multiple targets located at different range bins, we can filter out the spectrum of each target, and then implement above transformations for each target.

\subsection{Translating in Angle \label{apx_trans_ang}}

Implementing the Range FFT and Angle FFT on the digitized de-chirped signal \eqref{dechirp}, we obtain the 2D range-angle spectrum below:
\begin{equation}
\resizebox{0.435\textwidth}{!}{
 $ \begin{aligned}
  F_{\R{RA}}(m_{r}, m_\theta) = & \sum_{i=0}^{N_{\R{s}} - 1} \sum_{q=0}^{N_{\R{Rx}}-1}
    f(\frac{i}{f_{\R{s}}},q) \exp(-j\frac{2\pi m_{r} i}{M_r})
    \exp(-j\frac{2\pi m_\theta q}{M_\theta}) \\
    = & F_{\R{R}}(m_{r}) \sum_{q=0}^{N_{\R{Rx}}-1}  \exp(j2\pi \frac{q d\sin{\theta_1}}{\lambda})
  \exp(-j\frac{2\pi m_\theta q}{M_\theta}) \\
    = & F_{\R{R}}(m_{r}) \frac{1-\exp\left(j2\pi N_{\R{Rx}} \left(\frac{d\sin{\theta_1}}{\lambda} - \frac{m_\theta}{M_\theta}\right)\right)} {1-\exp \left(j2\pi \left(\frac{d\sin{\theta_1}}{\lambda}-\frac{m_\theta}{M_\theta} \right) \right)}
\end{aligned}$
}
\label{RAfft}
\end{equation}

\noindent where \resizebox{0.43\textwidth}{!}{$F_{\R{R}}(m_{r})=\sum\limits_{i=0}^{N_{\R{s}}-1} A_1  \exp \left(j2\pi \left(f_{\R{c}}\tau_1 + S \frac{i}{f_{\R{s}}} \tau_1 - \frac{1}{2}S \tau_1^2 \right) \right) \exp(-j\frac{2\pi m_{r} i}{M_r})$} is the Range FFT output for range bin $m_{r}$; $M_\theta$ is the number of points for Angle FFT; $m_\theta \in \{0, 1, ..., M_\theta-1\}$ is the angle bin index. The mapping relationship between any azimuth angle $\theta$ and bin index is:
\begin{equation}
    m_\theta=\lfloor \frac{M_\theta d \sin \theta}{\lambda} \rfloor
    \label{dftmap}
\end{equation}

If the target at $(r_1, \theta_1)$ has a translation in angle, $\Delta \theta$, as we defined in Section \ref{radar_aug}, i.e. $r^\prime_1=r_1, \theta^\prime_1=\theta_1 + \Delta \theta$, and the antenna gain ratio due to angle shift is $\frac{G_{\theta^\prime_1}}{G_\theta}$, then the corresponding FMCW de-chirped signal $f_{\R{new}}(t,q)$ for the target at new location $(r_1, \theta^\prime_1)$ is expressed as:
\begin{equation}
    \resizebox{0.435\textwidth}{!}{
    $f_{\R{new}}(t,q) = A_1 \frac{G_{\theta^\prime_1}}{G_\theta} \exp(j2\pi(f_{\R{c}}\tau_1 + S t \tau_1 - \frac{1}{2}S \tau_1^2 + \frac{qd\sin{\theta^\prime_1}}{\lambda}))$
    }
\end{equation}

Again, the range-angle spectrum for the digitized signal $f_{\R{new}}(\frac{i}{f_{s}},q)$ can be obtained with similar derivations in \eqref{RAfft} and \eqref{dftmap}. That is,
\begin{equation}
\resizebox{0.435\textwidth}{!}{
    $F_{\R{RA, new}}(m_{r}, m_\theta) = 
    \frac{G_{\theta^\prime_1}}{G_\theta}
  F_{\R{R}}(m_{r}) \frac{1-\exp\left(j2\pi N_{\R{Rx}} \left(\frac{d\sin{\theta^{\prime}_1}}{\lambda} - \frac{m_\theta}{M_\theta}\right)\right)} {1-\exp \left(j2\pi \left(\frac{d\sin{\theta^{\prime}_1}}{\lambda}-\frac{m_\theta}{M_\theta} \right) \right)}$
  }
  \label{new_ra_spectrum}
\end{equation}

With $\Delta m_{\theta_1} = \lfloor \frac{M_\theta d \left(\sin{\theta_1} - \sin{\theta^{\prime}_1} \right)}{\lambda} \rceil$, we can approximate new range-angle spectrum \eqref{new_ra_spectrum} with the old one \eqref{RAfft} by following relation:
\begin{equation}
    F_{\R{RA, new}}(m_{r}, m_\theta) \approx \frac{G_{\theta^\prime_1}}{G_\theta} F_{\R{RA}}(m_{r}, m_\theta + \Delta m_{\theta_1})
    \label{translate_angle_equation}
\end{equation}

Equation \eqref{translate_angle_equation} implies that we would obtain the new range-angle spectrum $F_{\R{RA, new}}(m_r, m_\theta)$ by shifting $\Delta m_{\theta_1}$ cells along the angular direction in the original spectrum $F_{\R{RA}}(m_r, m_\theta)$. Similarly, when there are multiple targets located at different angle bins, we can also filter out the spectrum of each target, and then implement above transformations for each target.

\end{document}